\documentclass[amsmath,amssymb,aps,floats,amsfonts,notitlepage,superscriptaddress,eqsecnum,nofootinbib,prd,a4paper,twocolumn]{revtex4-1}

\usepackage[]{graphicx}
\usepackage{hyperref}
\usepackage{url}
\usepackage{color}
\usepackage[usenames,dvipsnames,svgnames,table]{xcolor}
\usepackage{multirow}
\allowdisplaybreaks[1]

\newcommand{\Xin}{{X^{\text{in}}_{\ell \omega}}}
\newcommand{\Xup}{{X^{\text{up}}_{\ell \omega}}}
\newcommand{\XinPN}{{X^{\text{in(PN)}}_{\ell m}}}
\newcommand{\XupPN}{{X^{\text{up(PN)}}_{\ell m}}}
\newcommand{\XinMST}{{X^{\text{in(MST)}}_{\ell m}}}
\newcommand{\XupMST}{{X^{\text{up(MST)}}_{\ell m}}}

\newcommand{\mr}{q}

\begin{document}

\title{Analytical high-order post-Newtonian expansions for extreme mass ratio binaries}

\author{Chris Kavanagh}
\affiliation{School of Mathematical Sciences and Complex \& Adaptive Systems Laboratory, University College Dublin, Belfield, Dublin 4, Ireland.}

\author{Adrian C.~Ottewill}
\affiliation{School of Mathematical Sciences and Complex \& Adaptive Systems Laboratory, University College Dublin, Belfield, Dublin 4, Ireland.}

\author{Barry Wardell}
\affiliation{School of Mathematical Sciences and Complex \& Adaptive Systems Laboratory, University College Dublin, Belfield, Dublin 4, Ireland.}
\affiliation{Department of Astronomy, Cornell University, Ithaca, NY 14853, USA}

\date{\today}

\begin{abstract}
We present analytic computations of gauge invariant quantities for a point mass
in a circular orbit around a Schwarzschild black hole, giving results up to
15.5 post-Newtonian order in this paper and up to 21.5 post-Newtonian order in
an online repository. Our calculation is based on the functional series method
of Mano, Suzuki and Takasugi (MST) and a recent series of results by Bini and
Damour. We develop an optimised method for generating post-Newtonian expansions
of the MST series, enabling significantly faster computations. We also clarify
the structure of the expansions for large values of $\ell$, and in doing so
develop an efficient new method for generating the MST renormalised angular
momentum, $\nu$.
\end{abstract}

\maketitle

\section{Introduction}

The equations of motion of an extreme mass ratio binary system --- for example,
a solar mass black hole or neutron star in orbit around a massive black hole ---
are well approximated by a perturbative expansion of Einstein's equations. With
$\mr=m/M$ (the ratio of the smaller mass to the the larger one) as an expansion
parameter, at zero-th order in $q$ one obtains geodesic motion in the spacetime
of the larger mass. The order $q$ corrections are commonly referred to as
the first order self-force and are obtained by solving the linearized Einstein
equations around the background spacetime of the larger mass, $M$.

An alternative approach to the two-body problem --- valid when the constituents
are far apart --- is the post-Newtonian approximation, which expands the
Einstein equations in $v^2/c^2$, where $v$ is a representative velocity and
$c$ is the speed of light. In the context of binary systems, the post-Newtonian
expansion maps onto an expansion in $1/r$, where $r$ is the separation of the
two objects.

Recent years have seen synergistic development of the post-Newtonian and
perturbative approaches to the extreme mass ratio binary problem. Comparisons
of gauge invariant quantities computed in the two approaches have 
gone beyond simple (but valuable) crosschecks \cite{Blanchet:2009sd}. They have
enabled impressive
developments which would have been difficult, if not impossible, to achieve
in each theory on its own. For example, on the post-Newtonian end self-force
results have produced predictions of previously unknown terms in the
post-Newtonian expansion
\cite{Barack:2009ey,Damour:2009sm,Blanchet:2010cx,Blanchet:2010zd,Barack:2010ny,Dolan:2013roa,Dolan:2014pja}.
On the self-force end, comparisons with post-Newtonian calculations and with
Numerical Relativity simulations have given insight into the higher-order
perturbative corrections not included in the self-force approximation
\cite{LeTiec:2011bk,Bernuzzi:2014owa}.

Traditional approaches to the perturbative treatment of Einstein's equations
have relied on numerical solutions of differential equations as a means to
obtaining the linearized metric perturbation. More recently, alternative,
\emph{functional methods} have emerged as a compelling approach to obtaining
the linearized metric perturbation. In particular, there has been a surge of
interest in approaches pioneered by the work of Mano, Suzuki and Takasugi (MST)
\cite{Mano:1996mf},
in which one writes solutions of the linearized Einstein
equations in terms of a rapidly convergent series of hypergeometric functions.

Functional methods have the distinct advantage of giving a representation in 
terms of exact quantities, rather than a truncated numerical value. Since modern
computer algebra systems can efficiently evaluate hypergeometric functions to 
an essentially arbitrary number of digits, a numerical approach built on top
of functional methods is a powerful tool. Indeed, a parallel pair of works by
Shah and Pound \cite{Shah:2015nva}, and by Johnson-McDaniel, Shah and Whiting
\cite{Johnson-McDaniel:2015vva} has used
numerical functional methods to simultaneously obtain many of the same results
as we present here. Reassuringly, a direct comparison of our results with those
of \cite{Shah:2015nva,Johnson-McDaniel:2015vva} has shown the two to be in perfect agreement
\cite{ShahPrivate}, providing a strong independent check to both calculations.

In this work, we compute post-Newtonian expansions of the linearized metric
perturbation for a circular-orbit binary system, ignoring spin terms, up to
order $y^{21.5}$, where $1/y = (M \Omega)^{-2/3}$ is an invariant measure of
the radius defined through the orbital frequency, $\Omega$. From this, we
compute very high order post-Newtonian
approximations of all known gauge invariant quantities up to quadrupole order.
Our approach works in the Regge-Wheeler gauge (with a transformation to an
asymptotically flat gauge compatible with that used by post-Newtonian theory)
and is fundamentally based on
functional methods, but avoids any numerical evaluation (and the associated
numerical truncation). Instead, our method builds on a series of developments
by Bini and Damour \cite{Bini:2013zaa,Bini:2013rfa,Bini:2014ica,Bini:2014zxa}. Much of the calculation described here relies
heavily on the methods they developed, combined with some modifications which
allow the calculation to be efficiently taken to much higher post-Newtonian
order.

The layout of the paper is as follows. In Sec.~\ref{sec:methods} we give
details of our method, including an MST-based expansion for low $\ell$-multipole
modes and an exact large-$\ell$ approach for the higher multipole modes. In
Sec.~\ref{sec:metric-reconstruction} we give explicit expressions for the
metric perturbation in Regge-Wheeler gauge in terms of homogeneous solutions of
the Regge-Wheeler equation. In Sec.~\ref{sec:results} we summarise our results
and we conclude with a discussion in Sec.~\ref{sec:discussion}.
Finally, in Appendix~\ref{sec:gauge-invariants} we give equations
for the gauge invariant quantities we compute (written in terms of this metric
perturbation) and in Appendices~\ref{sec:DeltaU}, \ref{sec:DeltaPsi},
\ref{sec:DeltaLambda1}, \ref{sec:DeltaLambda2}, \ref{sec:DeltaLambda3} and
\ref{sec:DeltaLambdaB} we give our high-order post-Newtonian
expansion of the gauge invariant quantities. As the results become increasingly
unwieldy with increasing post-Newtonian order, we restrict our printed results
to order $y^{15.5}$ and opt instead to provide the higher-order terms
electronically \cite{online}.

This paper follows the conventions of Misner, Thorne and Wheeler
\cite{Misner:1974qy}; a ``mostly positive'' metric signature, $(-,+,+,+)$, is
used for the spacetime metric, the connection coefficients are defined by
$\Gamma^{\lambda}_{\mu\nu}=\frac{1}{2}g^{\lambda\sigma}(g_{\sigma\mu,\nu}
+g_{\sigma\nu,\mu}-g_{\mu\nu,\sigma}$), the Riemann tensor is
$R^{\alpha}{}_{\!\lambda\mu\nu}=\Gamma^{\alpha}_{\lambda\nu,\mu}
-\Gamma^{\alpha}_{\lambda\mu,\nu}
+\Gamma^{\alpha}_{\sigma\mu}\Gamma^{\sigma}_{\lambda\nu}
-\Gamma^{\alpha}_{\sigma\nu}\Gamma^{\sigma}_{\lambda\mu}$, the Ricci tensor and
scalar are $R_{\alpha\beta}=R^{\mu}{}_{\!\alpha\mu\beta}$ and
$R=R_{\alpha}{}^{\!\alpha}$, and the Einstein equations are
$G_{\alpha\beta}=R_{\alpha\beta}-\frac{1}{2}g_{\alpha\beta}R=8\pi
T_{\alpha\beta}$. Standard geometrized units are used, with $c=G=1$, but we
include the explicit dependence on $G$ and $c$ in our post-Newtonian expansions
in cases where they are convenient for post-Newtonian order counting. We use
the spherical coordinates $\{t,r,\theta,\phi\}$ for the background Schwarzschild
spacetime and write tensors in terms of these coordinate components.

\section{Homogeneous Solutions of the Regge-Wheeler equation}
\label{sec:methods}
We seek solutions to the homogeneous Regge-Wheeler equation,
\begin{align*}
\left(\frac{d^2}{dr_*^2}+\omega^2-V(r)\right)X_{\ell\omega}(r)=0
\end{align*}
where $V(r)=\left(1-\frac{2 M}{r}\right)\left(\frac{\ell(\ell+1)}{r^2}-\frac{6 M}{r^3}\right)$ and $d/dr_*=\left(1-\frac{2 M}{r}\right)d/dr$.
We assume much the same strategy as \cite{Bini:2013rfa} in approaching the
various $\ell$ mode regimes. Namely we use the analytical results of Zerilli \cite{Zerilli:1971wd}
with asymptotically flat gauge correction for $\ell=0,1$, generate the
analytical MST series solutions for `low' $\ell$ modes, and use a variation of
Bini \& Damour's post-Newtonian ansatz valid for large-$\ell$ values to tackle
the rest.

\subsection{Low $\ell$ solutions}

A collection of exact homogeneous solutions of the Regge-Wheeler equation are given in \cite{Mano:1996mf}. 
These are constructed via the introduction of a parameter $\nu$, the generalisation of  the angular number $\ell$ 
to non-integer values. This is used to build a formal perturbation of the Regge-Wheeler equation. 
The exact solutions are then written as series of hypergeometric ${}_2 F_1$ functions and irregular confluent 
hypergeometric $U$ functions, whose convergence is assured by the choice of $\nu$.

\subsubsection*{The MST expansion of  $\Xin(r)$}

Following a similar notation to \cite{Bini:2013rfa},   the horizon solution may be written as
\begin{align}
\Xin(r&)=C_{\text{(in)}}^\nu(x)\sum_{n=-\infty}^{\infty}a_n^\nu \times
\nonumber \\
& \overline{F}(n+\nu-1-i\epsilon,-n-\nu-2-i\epsilon,1-2i\epsilon,x) , \\
\Xin(r&\rightarrow r_{\text{h}})\rightarrow B^{\text{trans}}_{l \omega} e^{- i \omega r_{*}}, \nonumber
\end{align}
where $B^{\text{trans}}_{l \omega}$ can be determined analytically by the behaviour of $\Xin$ as $r\rightarrow r_{\text{h}}$,
and 
\begin{align*}
C_{\text{(in)}}^\nu(x)=e^{i \epsilon(x-1)}(-x)^{-i\epsilon}(1-x)^{-1},  \\
\overline{F}(a,b,c,\zeta)=\frac{\Gamma(a) \Gamma(b) }{\Gamma(c)} {}_2 F_1 (a,b,c;\zeta),  \\
x=1-\frac{c^2 r}{2 G M}, \qquad \qquad \epsilon=\frac{2 G M \omega}{c^3}.
\end{align*}
However, we find it more useful to instead work with 
the following decomposition:
\begin{align}
{}_2 & F_1(a,b,c;\zeta) =
\nonumber \\
&\, \frac{\Gamma(c) \Gamma(b-a)}{\Gamma(b) \Gamma(c-a)} (1-\zeta)^{-a} 
{}_2 F_1(a,c-b,a-b+1,\tfrac{1}{1-\zeta})
\nonumber \\
&\,+ \frac{\Gamma(c) \Gamma(a-b)}{\Gamma(a) \Gamma(c-b)} (1-\zeta)^{-b} {}_2 F_1(c-a,b,b-a+1,\tfrac{1}{1-\zeta}) 
\end{align}
Thus $\overline{F}\left(a,b,c,\zeta\right)=\overline{F}_1\left(a,b,c,\zeta\right)+\overline{F}_2\left(a,b,c,\zeta\right)$ with
\begin{align}
\label{eq:Fbar1}
\overline{F}_1 & \left(a,b,c,\zeta\right)=
\nonumber \\
& \frac{\Gamma(a) \Gamma(b-a)}{ \Gamma(c-a)} (1-\zeta)^{-a} {}_2 F_1(a,c-b,a-b+1,\tfrac{1}{1-\zeta})\\
\label{eq:Fbar2}
\overline{F}_2 & \left(a,b,c,\zeta\right)=
\nonumber \\
& \frac{\Gamma(b) \Gamma(a-b)}{ \Gamma(c-b)}(1-\zeta)^{-b} {}_2 F_1(c-a,b,b-a+1,\tfrac{1}{1-\zeta}) .
\end{align} 
As a note, this leads to the $\Xin=X_0^{\nu}+X_0^{-\nu-1}$ 
representation of Eq~(2.15) in \cite{Mano:1996mf} if we let $n\rightarrow-n$ in the $\overline{F}_1$ sum. Once again following \citep{Bini:2013rfa} we 
will keep track of the large-$r$, small-$\omega$ coupled expansions by using the small parameter $\eta=1/c$, and the Bini \& Damour variables $X_1=GM/r$, $X_2{}^{\!1/2}=\omega r$. 
In this way, each instance of  $X_1$ and $X_2$ must each come with an $\eta^2$.

At this point we can do an examination of the leading powers of $\eta$ against $n$ for the various terms in the sum. The structure 
of the MST series coefficients $a_n$ is described in Sec.~6 of \cite{Mano:1996mf}, but we present it again in terms of powers of $\eta$
and in a form (Table~\ref{tab:a}) that makes it straightforward to combine with other elements of the expansions. 
We note that the corresponding behaviour is also discussed
in the Living Review by Sasaki and Tagoshi \cite{Sasaki:2003xr} for the MST series coefficients for the Teukolsky function,
which simply differ from those of the spin-2 Regge-Wheeler equation by a complex conjugation~\cite{CO:15}. 

\begin{table*}[htb]
\begin{center}
 \begin{tabular}{|c|c|c|c|c|c|c|}
\hline
& $n\leq -2\ell-1$ & $-2\ell \leq n\leq -\ell-1$ & $n=-\ell$ & $n=-\ell+1$ &  $n\geq -\ell+2$  \\
\hline
\hline
$a_n$ &$3(|n|-1)$&$3(|n|+1)$ &  $3\ell+6$ &  $3(\ell-1)+6$&  $3|n|$ \\
\hline
\end{tabular}
\caption{The leading behaviour of the MST coefficients for the spin-2 Regge-Wheeler equation in terms of powers of $\eta$. 
For example, $a_n= O(\eta^{3|n|})$ for $n\geq -\ell+2$.}
\label{tab:a}
\end{center}
\end{table*}

The structure of the series coefficients is not the only irregular behaviour we need to account for; the dependence on $\eta$ of both the parameters and argument of 
the ${}_2 F_1$ also go through order jumps, which are tabulated in Table~\ref{tab:F}. In constructing this and subsequent Tables
the key elements are that
\begin{align*}
\epsilon\sim \eta^3, \qquad \Delta\nu\equiv \nu - \ell \sim \epsilon^2 \sim \eta^6, \qquad \frac{1}{1-x} \sim \eta^2 .
\end{align*}
The scaling of $\epsilon$ and $x$ with $\eta$ come from their definitions,
whereas the scaling of $\nu$ can be seen from its low-frequency expansion.  

One subtlety here as that while we tend to think of ${}_2 F_1$ as being regular in its argument, this is only true if the
parameters are held fixed. It is easy to see, for example, that the expansion of the hypergeometric function in 
Eq.~(\ref{eq:Fbar1}) is given by
\begin{align*}
1 + \tfrac{(n+\ell-1-i \epsilon +\Delta\nu)(n+\ell+3-i \epsilon +\Delta\nu)}{2(n+\ell+1-\Delta\nu)}  X_1 \eta^2+ \cdots ,
\end{align*}
and the denominator is manifestly $\eta^6$ when $n+\ell+1=0$, leading to a
divergent limit as $\eta\to0$. This divergence is more than compensated for by
the behaviour of the prefactor, but leads to larger term than a casual analysis
would suggest. This term corresponds to the daggered entries in
Table~\ref{tab:F}.

\begin{table*}[htb]
\begin{center}
 \begin{tabular}{|c|c|c|c|c|c|c|}
\hline
& $n\leq -\ell-3$ & $n=-\ell-2$& $n= -\ell-1$ & $n=-\ell$ & $n=-\ell+1$ &  $n\geq -\ell+2$  \\
\hline
\hline
$\overline{F}_1$&$2n+2\ell-5$&$-9$&$-19^\dagger$&$-11$&$-9$&$2n+2\ell-5$\\
$\overline{F}_2$&$-2n-2\ell-7$&$-9$&$-11$&$-11^\dagger$&$-9$&$-2n-2\ell-7$\\
\hline
\end{tabular}
\caption{The leading behaviour of the hypergeometric functions appearing in Eqs.~(\ref{eq:Fbar1}) and (\ref{eq:Fbar2}). The terms marked with a $^\dagger$ highlight the interplay between parameters and argument; see the text for a full discussion.}
\label{tab:F}
 \end{center}
\end{table*}

\begin{table*}[htb]
\begin{center}
\hspace{-1cm} 
\begin{tabular}{|c|c|c|c|c|c|c|c|}
\hline
& $n\leq -2\ell-1 $&$-2\ell\leq n\leq -\ell-3$ & $n=-\ell-2$  &  $n= -\ell-1$ & $n=-\ell$ & $n=-\ell+1$ &  $n\geq -\ell+2$  \\
\hline
\hline
$\eta^{2\ell+7}a_n\overline{F}_1$&$|n|+4l-1$&$|n|+4\ell+5$&$6\ell+7$&$5\ell-6^\dagger$&$5\ell+2$&$5\ell+1$&$3|n|+2n+4l+2$\\
$\eta^{2\ell+7}a_n\overline{F}_2$&$5|n|-3$&$5|n|+3$&$6\ell+7$&$5\ell+2$&$5\ell+2^\dagger$&$5\ell+1$&$3|n|-2n$\\
\hline
\end{tabular}
 \end{center}
\caption{The combined behaviour following from Tables~\ref{tab:a} and \ref{tab:F}. We take out a factor of $\eta^{2l+7}$
for normalisation fixing the largest term as $O(1)$,  everything higher can be read as relative.}
\label{tab:aF}
\end{table*}

Combining the behaviour of the various elements gives a master table, Table~\ref{tab:aF}, for the orders of each term in the series. 
This turns out to be the most useful table for optimising our calculations of the inner
 solution since they allow us to say with certainty what we do and do not need to calculate to
  correctly give a desired order. For example, one can see immediately from looking 
  at these that for $n>0$ our contribution from $\overline{F}_1$ will die out quickly and
   can be largely ignored. 

\subsubsection*{The MST expansion of  $\Xup(r)$}

The solution satisfying the boundary condition at infinity is
\begin{align}
X_{\ell\omega}^{{\rm up}} (r&) = C_{({\rm up})}^{\nu} (z) \sum_{n=-\infty}^{+\infty} a_n^{\nu} (-2iz)^n \times
\nonumber \\
& \, \overline{U} (n+\nu+1 -i\epsilon, 2n+2\nu+2,-2iz), \\
\Xup(r&\rightarrow \infty)\rightarrow C^{\text{trans}}_{l \omega} e^{ i \omega r_{*}}, \nonumber
\end{align}
where $C^{\text{trans}}_{l \omega}$ can be determined analytically by the behaviour of $\Xup$ as $r\rightarrow \infty$,
and 
\begin{align}
C_{({\rm up})}^{\nu} (z)=&e^{iz}z^{\nu+1}(1-\frac{\epsilon}{z})^{-i\epsilon}2^{\nu}e^{-\pi \epsilon}e^{-i\pi(\nu+1)}  \\
\overline{U}(a,b,\zeta) &= \frac{\Gamma (a-2) \Gamma (a)}{\Gamma (a^*+2)\Gamma (a^*) }U(a,b,\zeta)
\end{align}
and $z=\frac{\omega r}{c}$. In a similar vein to the inner case we split this solution by making use of the identity 
\begin{align}
\mathop{U\/}\nolimits\!\left(a,b,z\right)=& \frac{\Gamma\left(1-b\right)}{\Gamma\left(a-b+1\right)}M\left(a,b,z\right)
\nonumber \\
& \quad
+\frac{\Gamma\left(b-1\right)}{\Gamma\left(a\right)}z^{1-b}M\left(a-b+1,2-b,z\right).
\end{align}
Thus $\overline{U}\left(a,b,\zeta\right)=\overline{U}_1\left(a,b,\zeta\right)+\overline{U}_2\left(a,b,\zeta\right)$
\begin{align}
\label{eq:Ubar1}
\overline{U}_1\left(a,b,\zeta\right)=&\frac{\Gamma (a-2) \Gamma (a)\Gamma\left(1-b\right)}{\Gamma (a^*+2)\Gamma (a^*)\Gamma\left(a-b+1\right)}M\left(a,b,\zeta\right)\\
\label{eq:Ubar2}
\overline{U}_2\left(a,b,\zeta\right)=& \frac{\Gamma (a-2)\Gamma\left(b-1\right)}{\Gamma (a^*+2)\Gamma (a^*)}z^{1-b} \times 
\nonumber \\
& \qquad M\left(a-b+1,2-b,\zeta\right).
\end{align}

\begin{table*}[htb]
\begin{center}
 \begin{tabular}{|c|c|c|c|c|c|c|}
\hline
& $n\leq -\ell-3$ & $n=-\ell-2$& $n= -\ell-1$ & $n=-\ell$ & $n=-\ell+1$ &  $n\geq -\ell+2$  \\
\hline
\hline
$\overline{U}_1$&$0$&$-3$&$-5^\dagger$&$-6$&$-6$&$-3$\\
$\overline{U}_2$&$-2n-2\ell-4$&$-3$&$-5$&$-6^\dagger$&$-6$&$-2n-2\ell-1$\\
\hline
\end{tabular}
\caption{The leading behaviour of the hypergeometric functions appearing in Eqs.~(\ref{eq:Ubar1}) and (\ref{eq:Ubar2}).The terms marked with a $^\dagger$ highlight the interplay between parameters and argument in this case. }
\label{tab:U}
 \end{center}
\end{table*}

\begin{table*}[htb]
\begin{center}
\hspace{-1cm} 
\begin{tabular}{|c|c|c|c|c|c|c|c|}
\hline
& $n\leq -2\ell-1 $&$-2\ell\leq n\leq -\ell-3$ & $n=-\ell-2$  &  $n= -\ell-1$ & $n=-\ell$ & $n=-\ell+1$ &  $n\geq -\ell+2$\\
\hline
\hline
$\eta^{2\ell+1}a_n(-2iz)^n\overline{U}_1$&$2|n|+2\ell-2$&$2|n|+2\ell+4$&$4\ell+5$&$4\ell+1^\dagger$&$4\ell+1$&$4\ell-1$&$3|n|+n+2\ell-2$\\
$\eta^{2\ell+1}a_n(-2iz)^n\overline{U}_2$&$4|n|-6$&$4|n|$&$4\ell+5$&$4\ell+1$&$4\ell+1^\dagger$&$4\ell-1$&$3|n|-n$\\
\hline
\end{tabular}
\end{center}
\caption{The combined behaviour following from Tables~\ref{tab:a} and \ref{tab:U}.}
\label{tab:aU}
\end{table*}

Analysis of the behaviour of the $M$ functions for small $\eta$ produces Table~\ref{tab:U}.
Again we must take due care that parameters as well as the argument depends on $\eta$.  For example, when considering $\overline{U}_1\left(n+\ell+1+\Delta\nu -i\epsilon, 2(n+\ell+1)+2\Delta\nu,-2iz\right)$ the Taylor 
coefficients in the expansion of $M$ are regular if $n+\ell\geq0$  but if $n+\ell=-k$ $(k \in \mathbb{N})$ the Taylor coefficients of order $(-2iz)^{2k-1}$ 
and beyond are order $\eta^{-3}$.  This means that for $k=1$ the function behaves as $\eta^{-2}$, while for $k\geq 2$
the behaviour is regular but we need to include 3 more terms than one might have expected to get to the appropriate 
$\eta$ order.
Correspondingly when considering $\overline{U}_2\left(n+\ell+1+\Delta\nu -i\epsilon, 2(n+\ell+1)+2\Delta\nu,-2iz\right)$ the Taylor 
coefficients in the expansion of $M$ are regular if $n+\ell<0$  but if $n+\ell=k$ $(k \in \mathbb{N}\cup\{0\})$ the Taylor coefficients of order $(-2iz)^{2k+1}$ 
and beyond are order $\eta^{-3}$.  This means that for $k=0$ the function behaves as $\eta^{-2}$, while for $k\geq 1$
the behaviour is regular but we need to include 3 more terms than one might have expected to get to the appropriate 
$\eta$ order.

Combining the behaviour of the various elements gives a master table, Table~\ref{tab:aU}, for the orders of each term in the series for $X_{\text{up}}$.

\subsection{Large $\ell$ solutions}

As described in \citep{Bini:2013rfa}, one can generate `large-$\ell$'
 homogeneous solutions using the template
\begin{align}
\XinPN =&r^{\ell+1}\big[1+\eta^2 A_2^{PN,\ell}+\eta^4 A_4^{PN,\ell}
\nonumber \\
& \qquad \qquad +\eta^6 A_6^{PN,\ell}+\ldots\big]  \nonumber \\
\XupPN =&r^{-\ell}\big[1+\eta^2 A_2^{PN,-\ell-1}+\eta^4 A_4^{PN,-\ell-1}
\nonumber \\
& \qquad \qquad +\eta^6 A_6^{PN,-\ell-1}+\ldots\big]    \label{BD:PN}
\end{align}
where the $A_i$ are for the most part polynomials in $X_1$ and $X_2{}^{\!1/2}$, but one 
finds extra $r$ dependent $\log$ terms appearing after the sixth order, $\log^2$ terms appearing after the twelfth order, etc.
However, the precise details appear opaque.

In generating the MST expansions for high values of $\ell$ and for general $\ell$ above the given order
the full structure of this expansion becomes clear:

\vspace{1.5mm}
\noindent~(1) In the PN expansion of the MST series using the Bini-Damour variables $X_1$, $X_2$ the combination
\begin{align*}
\epsilon=2 X_1 X_2{}^{\!1/2}\eta^3
\end{align*}
is $r$-independent.  A phase of the form
\begin{align*}
\psi^\text{in/up}=  \sum_{i=0}^\infty \psi^\text{in/up}_i \epsilon^i = \sum_{i=0}^\infty \psi^\text{in/up}_i (2 X_1 X_2{}^{\!1/2}\eta^3)^i
\end{align*}
is therefore irrelevant as it amounts to a normalisation constant, which can
be seen to cancel upon division by the Wronskian in
Sec.~\ref{sec:metric-reconstruction}.
\vspace{1.5mm}

\noindent~(2) Having factored out an $r$-independent phase of the type described in~(1), we find the MST expansions for $X^\text{in}$ may be further expressed in the form 
\begin{align}
\label{eq:in_ansatz}
\XinMST= e^{i \psi^\text{in}} X_1{}^{-\ell - 1 - 
 \sum\limits_{j=1}^{\infty} a_{(6 j, 2 j)} (2 X_1 X_2{}^{\!1/2} \eta^3)^{2 j}} \times
\nonumber \\
 \left[1+\eta^2 A_2^{\ell}+\eta^4 A_4^{\ell}+\eta^6 A_6^{\ell}+\ldots\right]
\end{align}
where the $A_i$ are now \emph{strictly} polynomials in $X_1$, $X_2$ which we may express as
\begin{align*}
A_{2n}^\ell=& \sideset{}{'}\sum\limits_{i=0}^{n}  a_{(2n,i)} X_1{}^i  (X_2{}^{\!1/2})^{2(n-i)}
\nonumber \\
=& \sideset{}{'}\sum\limits_{i=0}^{n}  a_{(2n,i)} X_1{}^i  X_2{}^{\! n-i},
\end{align*}
where the prime indicates that we omit terms where the combination is a power of $X_1 X_2{}^{\!1/2}$ as these are already included in the phase term. This occurs if $i =  2(n-i)$ which is only possible if $2n = 6j$ and $i=2j$ for some $j\in\mathbb{N}$, explaining our notation 
for the phase expansion.

\begin{figure}[htb!]
\includegraphics[scale=0.51]{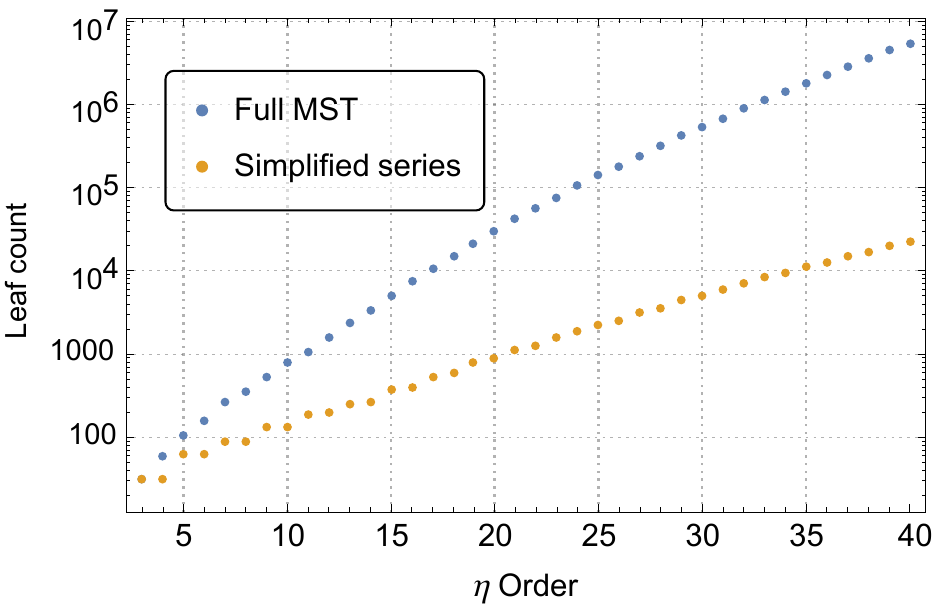}
\includegraphics[scale=0.5]{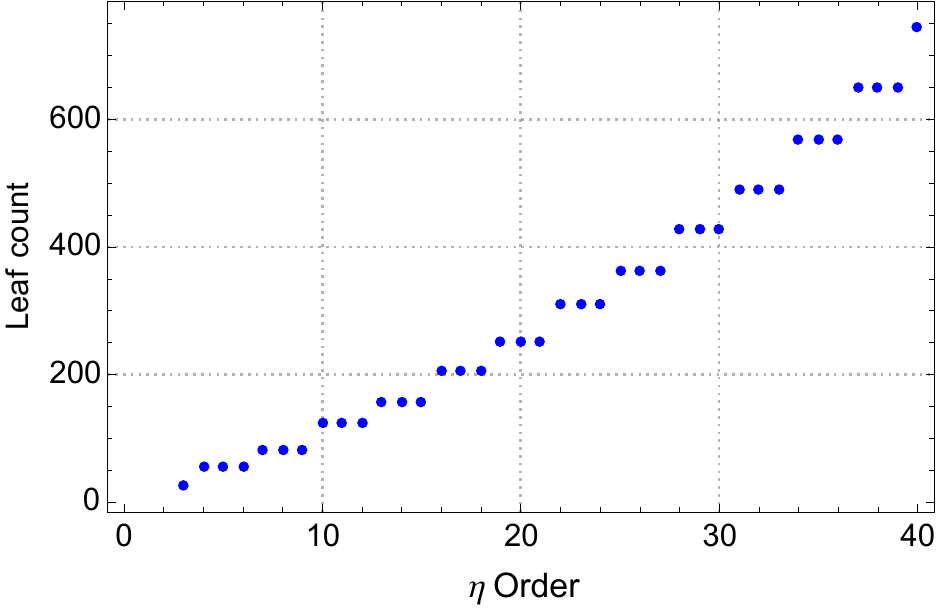}
\caption{A demonstration of the algebraic simplification obtained by rewriting the MST series for $\Xup$
 as phase$\times$log$\times$PN series for $\ell=5$. Plots use the
Mathematica LeafCount function to measure the expression complexity with increasing $\eta$-order. 
The first figure shows the full MST expansion 
along with the the remaining `PN series' on a log scale, and the second shows the phase plus log terms. Note that we can mostly ignore the phase and log terms as they 
will drop out when dividing by the Wronskian.}
\label{fig:Leafcounts}
\end{figure}

In our calculations in the previous subsection we calculated $X_{\text{in }}$
from the hypergeometric series to order $\eta^{40}$ up to $\ell=20$ explicitly, and then rewrote
it in the form of Eq.~\eqref{eq:in_ansatz} prior to calculating the metric
perturbation, as it offered immense simplifications of the algebra (see Fig.~\ref{fig:Leafcounts}).

However, for large-$\ell$
we may take an alternative approach. Acting on a function
$f(X_1,X_2{}^{\!1/2})$, we have
\begin{align*}
r \frac{\text{d}f}{\text{d} r} = -X_1 \frac{\text{d}f}{\text{d} X_1}  + X_2{}^{\!1/2}  \frac{\text{d}f}{\text{d} X_2{}^{\!1/2}},
\end{align*}
and use this substitution in the spin-$s$ Regge-Wheeler operator, which is
given by
\begin{align*}
 (1 & - 2 X_1 \eta^2)^2 r \frac{\text{d}}{\text{d} r} \left[
    r \frac{\text{d}\ }{\text{d} r} \right]  + (1 - 2 X_1 \eta^2) (4 X_1 \eta^2 - 1)  r \frac{\text{d}}{\text{d} r}
\nonumber \\
& + \eta^2 X_2 - (1 - 
     2 X_1 \eta^2) \bigl(\ell (\ell + 1) + (1 - s^2) 2 X_1 \eta^2\bigr) 
\end{align*}
Requiring that the ansatz of \eqref{eq:in_ansatz} satisfy the spin-2 Regge-Wheeler
equation immediately determines all coefficients at orders up to $n=2\ell$. This
means that we may determine at any given order the
necessary expansion without solving the full MST expansions, as Bini \& Damour did
in the context of \ref{BD:PN}. (As a check we have
solved the MST expansion for large $\ell$ explicitly to order $\eta^{21}$ and
found full agreement with the much simpler method described here.)

An additional bonus of this method is that one may identify the coefficients
$a_{(6 j, 2 j)}$ as being precisely the coefficients of the renormalised
angular momentum $\nu_-$ corresponding to $-l-1$, that is 
\begin{align}
 \nu_-(\epsilon) = -\ell-1 - \sum\limits_{j=1}^{\infty} a_{(6 j, 2 j)} \epsilon^{2 j}
\end{align}
or, in other words, we have
\begin{align}
\XinMST=& e^{i \psi^\text{in}} X_1{}^{\nu_-(2 X_1 X_2{}^{\!1/2} \eta^3)} \times
\nonumber \\
& \quad \left[1+\eta^2 A_2^{\ell}+\eta^4 A_4^{\ell}+\eta^6 A_6^{\ell}+\ldots\right] .
\end{align}
While we discovered this behaviour by an explicit perturbative expansion, in
hindsight, the appearance of $\nu$ in this equation might have been expected on
monodromy grounds \cite{Castro:2013lba}.

\noindent~(3) The full MST expansions for $X^\text{up}$ are in general much more complicated, involving, for example, high powers of $\log X_2$. Nonetheless, the leading terms are again easy; with the ansatz
\begin{align}
\label{eq:up_ansatz}
\XupMST&= e^{i \psi^\text{up}} (X_2{}^{\!1/2}){}^{-l - 
 \sum\limits_{j=1}^{\infty} b_{(6 j, 2 j)} (2 X_1 X_2{}^{\!1/2} \eta^3)^{2 j}} \times
\nonumber \\
& \left[1+\eta^2 B_2^{\ell}+\eta^4 B_4^{\ell}+\dots+\eta^{2\ell} B_{2\ell}^{\ell}+O(\eta^{2\ell+2})\right] ,
\end{align}
the  Regge-Wheeler equation again determines all the given coefficients (that is, orders up to $n=2l$)
and 
\begin{align}
 \nu_+(\epsilon) = \ell + \sum\limits_{j=1}^{\infty} b_{(6 j, 2 j)} \epsilon^{2 j} .
\end{align}

In both cases it is easy to see why we can determine to this order and no higher since the $r$ independent factor
$(2 X_1 X_2{}^{\!1/2}\eta^3)^{\pm(2l+1)}$ will serve to mix them at that order.  To obtain the correct solutions we must
supplement the Regge-Wheeler equation with the boundary condition; our simple ansatz cannot implement these while 
the MST expansion does.

\section{Reconstructed metric perturbation in Regge-Wheeler gauge}
\label{sec:metric-reconstruction}
In Regge-Wheeler gauge, the components of the metric perturbation may be
written in terms of the homogeneous solutions, $\Xin$ and $\Xup$. For the case
of a point mass on a circular orbit of radius $r_0$, the metric perturbation
components are given explicitly below for the case $r<r_0$ (we ignore terms
involving $\delta(r-r_0)$ and its derivatives since they are not relevant to
our calculations). Here, $W=\Xin(r_0) \Xup'(r_0) - \Xup(r_0) \Xin'(r_0)$ is the
Wronskian, $\lambda \equiv (\ell-1)(\ell+2)/2$ and $\Lambda \equiv \lambda + 3
M / r$. To obtain the components for $r>r_0$, simply interchange $\Xin$ and
$\Xup$ in these expressions.

The $(t,t)$ component of the metric perturbation is given by
\begin{widetext}
\begin{equation}
h^{\rm RW}_{tt} = \frac{\pi Y^\ast_{\ell m}(\theta_0, \phi_0) Y_{\ell m}(\theta, \phi) (h_{tt}^{\text{in}} \Xin(r) + h_{tt}^{\text{in}'} \Xin'(r)) (h_{tt}^{\text{up}} \Xup(r_0) + h_{tt}^{\text{up}'} \Xup'(r_0))}{8 \lambda (\lambda +1) r^3 r_0^{5/2} W (r_0 - 2 M) (r_0-3 M)^{1/2} (9 m^2 M^3+\lambda ^2 (\lambda +1)^2 r_0^3)},
\end{equation}
where
\begin{align}
  h_{tt}^{\text{in}} =& (6 M-2 (\lambda +1) r) (4 \Lambda m^2 M r^3+4 \lambda (\lambda +1) r_0^3 (2M-r)), \nonumber \\
  h_{tt}^{\text{in}'} =& 2 r (2 M-r) (12 m^2 M^2 r^2+4 \lambda (\lambda +1) r_0^3 (3 M-r)), \nonumber \\
  h_{tt}^{\text{up}} =& -8 (6 M^3 (-\lambda +m^2-1)+3 M^2 r_0 (2 \lambda ^2+3 \lambda -m^2+1)-\lambda (\lambda +1) M r_0^2 (\lambda +m^2+4)+\lambda (\lambda +1)^2 r_0^3), \nonumber \\
  h_{tt}^{\text{up}'} =& 8 r_0 (2 M-r_0) ((\lambda +1) (3 M^2+\lambda r_0 (r_0-2 M))-3 m^2 M^2).
\end{align}
The $(t,r)$ component of the metric perturbation is given by
\begin{equation}
h^{\rm RW}_{tr} =  -\sqrt{\frac{M}{r_0-3 M}} \frac{i \pi m Y^\ast_{\ell m}(\theta_0, \phi_0) Y_{\ell m}(\theta, \phi) (h_{tr}^{\text{in}} \Xin(r) + h_{tr}^{\text{in}'} \Xin'(r)) (h_{tr}^{\text{up}} \Xup(r_0) + h_{tr}^{\text{up}'} \Xup'(r_0))}{4 \lambda (\lambda +1) r^2 r_0^4 W (2 M-r) (2 M-r_0) (9 m^2 M^3+\lambda
 ^2 (\lambda +1)^2 r_0^3)},
\end{equation}
where
\begin{align}
  h_{tr}^{\text{in}} =& 12 m^2 M^2 r^4-4 r_0^3 (18 M^3-3 (2 \lambda +5) M^2 r-3 (\lambda ^2-1) M r^2+\lambda (\lambda +1) r^3), \nonumber \\
  h_{tr}^{\text{in}'} =& -4 \Lambda r^2 r_0^3 (r-2 M) (-3 M+\lambda r+r), \nonumber \\
  h_{tr}^{\text{up}} =&  8 (6 M^3 (-\lambda +m^2-1)+3 M^2 r_0 (2 \lambda ^2+3 \lambda -m^2+1)-\lambda (\lambda +1) M r_0^2 (\lambda +m^2+4)+\lambda (\lambda +1)^2 r_0^3), \nonumber \\
  h_{tr}^{\text{up}'} =& -8 r_0 (2 M-r_0) ((\lambda +1) (3 M^2+\lambda r_0 (r_0-2 M))-3 m^2 M^2).
\end{align}
The $(r,r)$ component of the metric perturbation is given by
\begin{equation}
h^{\rm RW}_{rr} = -\frac{\pi Y^\ast_{\ell m}(\theta_0, \phi_0) Y_{\ell m}(\theta, \phi) (h_{rr}^{\text{in}} \Xin(r) + h_{rr}^{\text{in}'} \Xin'(r)) (h_{rr}^{\text{up}} \Xup(r_0) + h_{rr}^{\text{up}'} \Xup'(r_0))}{8 \lambda (\lambda +1) r r_0^{5/2} W (r-2 M)^2 (2 M-r_0) \sqrt{r_0-3 M} (9 m^2 M^3+\lambda ^2 (\lambda +1)^2
 r_0^3)},
\end{equation}
where
\begin{align}
  h_{rr}^{\text{in}} =& (6 M-2 (\lambda +1) r) (4 \Lambda m^2 M r^3+4 \lambda (\lambda +1) r_0^3 (2 M-r)), \nonumber \\
  h_{rr}^{\text{in}'} =& 2 r (2 M-r) (12 m^2 M^2 r^2+4 \lambda (\lambda +1) r_0^3 (3 M-r)), \nonumber \\
  h_{rr}^{\text{up}} =& -8 (6 M^3 (-\lambda +m^2-1)+3 M^2 r_0 (2 \lambda ^2+3 \lambda -m^2+1)-\lambda (\lambda +1) M r_0^2 (\lambda +m^2+4)+\lambda (\lambda +1)^2 r_0^3), \nonumber \\
  h_{rr}^{\text{up}'} =& 8 r_0 (2 M-r_0) ((\lambda +1) (3 M^2+\lambda r_0 (r_0-2 M))-3 m^2 M^2).
\end{align}
The $(t,\theta)$ and $(t,\phi)$ components of the metric perturbation are given by
\begin{equation}
h^{\rm RW}_{t\theta} = - i m H_t \csc \theta Y_{\ell m}(\theta, \phi), \quad
h^{\rm RW}_{t\phi} = H_t \sin \theta Y_{\ell m,\theta}(\theta, \phi),
\end{equation}
where
\begin{equation}
H_t = \sqrt{\frac{M}{r_0-3 M}} \frac{4 \pi Y^\ast_{\ell m,\theta}(\theta_0, \phi_0) (2 M-r) (r \Xin'(r)+\Xin(r)) (r_0
 \Xup'(r_0)+\Xup(r_0))}{\lambda (\lambda +1)\, r\, r_0 W}.
\end{equation}
The $(r,\theta)$ and $(r,\phi)$ components of the metric perturbation are given by
\begin{equation}
h^{\rm RW}_{r\theta} = - i m H_r \csc \theta Y_{\ell m}(\theta, \phi), \quad
h^{\rm RW}_{r\phi} = H_r \sin \theta Y_{\ell m,\theta}(\theta, \phi),
\end{equation}
where
\begin{equation}
H_r = -\frac{4 i \pi Y^\ast_{\ell m,\theta}(\theta_0, \phi_0) m M r^2 \Xin(r) (r_0 \Xup'(r_0)+\Xup(r_0))}{\lambda (\lambda +1)
 r_0^{5/2} W (2 M-r) (r_0-3 M)^{1/2}}.
\end{equation}
The $(\theta,\theta)$ and $(\phi,\phi)$ components of the metric perturbation are given by
\begin{equation}
  h^{\rm RW}_{\theta\theta} = K, \quad
  h^{\rm RW}_{\phi\phi} = \sin^2 \theta K,
\end{equation}
where
\begin{equation}
K = \frac{\pi Y^\ast_{\ell m}(\theta_0, \phi_0) Y_{\ell m}(\theta, \phi) (K^{\text{in}} \Xin(r) + K^{\text{in}'} \Xin'(r)) (K^{\text{up}} \Xup(r_0) + K^{\text{up}'} \Xup'(r_0))}{8 \lambda (\lambda +1) \,r\, r_0^{5/2} W (r_0-2 M) (r_0-3 M)^{1/2} (9 m^2 M^3+\lambda ^2 (\lambda +1)^2
 r_0^3)},
\end{equation}
with
\begin{align}
  K^{\text{in}} =& 24 m^2 M^2 r^3-8 (\lambda +1) r_0^3 (-6 M^2+3 M r+\lambda (\lambda +1) r^2), \nonumber \\
  K^{\text{in}'} =& -8 (\lambda +1) r r_0^3 (r-2 M) (3 M+\lambda r), \nonumber \\
  K^{\text{up}} =& 8 (6 M^3 (-\lambda +m^2-1)+3 M^2 r_0 (2 \lambda ^2+3 \lambda -m^2+1)-\lambda (\lambda +1) M r_0^2 (\lambda +m^2+4)+\lambda (\lambda +1)^2 r_0^3), \nonumber \\
  K^{\text{up}'} =& -8 r_0 (2 M-r_0) ((\lambda +1) (3 M^2+\lambda r_0 (r_0-2 M))-3 m^2 M^2).
\end{align}
The $(\theta, \phi)$ component is exactly zero, as required by the Regge-Wheeler
gauge conditions.
\end{widetext}

The above expressions may only be used for the multipole modes $\ell \ge 2$.
For completeness, we also give the metric perturbation components for $\ell=0,1$, which were
derived analytically by Zerilli \cite{Zerilli:1971wd}. As in Ref.~\cite{Bini:2013rfa}, a correction term
is added to Zerilli's solutions to account for the change to an
asymptotically flat gauge compatible with the one used in post-Newtonian theory:
\begin{align}
h_{tt}^{\ell=0}&=\frac{2\tilde{E_1}}{r_0}\frac{1-\frac{2M}{r}}{1-\frac{2M}{r_0}}H(r_0-r)+\frac{2\tilde{E_1}}{r}H(r-r_0), \nonumber \\
h_{rr}^{\ell=0}&=\frac{2\tilde{E_1}}{r\left(1-\frac{2M}{r}\right)^2}H(r-r_0),
\end{align}
with $\tilde{E_1}=\frac{1-2 M/r_{0}}{ \sqrt{1-3 M/r_0}}$. For $\ell=1$ we have a contribution from the odd sector,
\begin{align}
h_{t\phi}^{\ell=1}&=-2\tilde{L}_1 \sin^2\theta\left[\frac{r^2}{r_0^3}H(r_0-r)+\frac{1}{r}H(r-r_0)\right],
\end{align}
and a contribution from the even sector,
\begin{align}
h_{tt}^{\ell=1}&=-2\tilde{E}\frac{r_0-2M}{r(r-2M)}\left[1-\frac{r^3\Omega}{M}\right]\sin\theta\cos\overline{\phi}H(r_0-r), \nonumber \\
h_{tr}^{\ell=1}&=6\tilde{E}\Omega\frac{r(r_0-2M)}{(r-2M)^2}\sin\theta\sin\overline{\phi}H(r_0-r), \nonumber \\
h_{rr}^{\ell=1}&=-6\tilde{E}\frac{r(r_0-2M)}{(r-2M)^3}\sin\theta\cos\overline{\phi}H(r_0-r),
\end{align}
where $H$ is the Heaviside function and $\tilde{L}_1=\sqrt{\frac{r_0^2}{M(r_0-3M)}}$.

\section{Regularized Calculation of Gauge-invariant Quantities}

With the spherical-harmonic modes of the Regge-Wheeler metric perturbation at
hand, the final stage in our calculation is to compute the known gauge invariant
quantities. To do so, we use the expressions given in
Appendix~\ref{sec:gauge-invariants} for the gauge invariants in terms of the
metric perturbation, and then sum over $m$ (which is straightforward) and over
$\ell$ (which deserve further attention and will be addressed in detail next).

We have so far only discussed the computation of the retarded metric
perturbation, yet the retarded perturbation diverges on the worldline - this
divergence manifests itself through the failure of the sum over $\ell$ to
converge. Essentially, we address the issue of regularization of this
divergence using the standard mode-sum regularization approach
\cite{Barack:2002bt}. More explicitly, we compute PN expansions of the (tensor)
spherical harmonic modes of each invariant using the retarded Regge-Wheeler
metric perturbation, and then subtract a sufficient number of PN-expanded
regularization parameters such that the sum over modes yields a finite,
convergent result. This produces invariants of the perturbed spacetime defined
by the Detweiler-Whiting regular field \cite{Detweiler:2002mi}, not of the full
retarded-field metric.

It turns out that no new subtleties appear for our circular-orbit case in the
Regge-Wheeler gauge, as compared to the standard Lorenz gauge approach. A full
analysis is beyond the scope of this paper and will instead be addressed
through a more detailed analysis in a future work. We do
note, however, that it is perhaps not surprising given that all of the
quantities we are computing are explicitly invariant under gauge
transformations represented by a gauge vector which respects the helical
symmetry of the problem, i.e. $\xi^\phi = \Omega \xi^{t}$. Furthermore, in
Ref.~\cite{Dolan:2014pja} we computed with others highly accurate numerical
values for the invariants in both Lorenz and Regge-Wheeler gauges and found
that the individual retarded field (tensor) spherical harmonic modes are in
perfect agreement between the two gauges, to within the highly-constrained
error bars of our numerical results. We also used the standard (and robust)
procedure of projecting the Lorenz-gauge modes onto scalar spherical harmonics,
combined with analytically derived regularization parameters (which were
obtained using the Detweiler-Whiting singular metric perturbation) as a
validation of our results.

In practice, we found it convenient to take some short-cuts in our
calculation, which are well-justified by our validation and consistency
checks. Specifically, the approach we take is as follows:
\begin{enumerate}
\item For each invariant, we compute the retarded field tensor harmonic
   $\ell$-modes (summed over $m$) using the retarded metric perturbation on
   both (radial) sides of the worldline, i.e. by interchanging $X_{in}$ and
   $X_{up}$ in the expressions for the metric perturbation.
\item We eliminate odd divergent powers of $\ell$ (i.e. the parts proportional to
   $2\ell+1$ and $\ell(\ell+1)(2\ell+1)$) by averaging the results on either
   side of the worldline.
\item We make use of the post-Newtonian expansion of the first two Lorenz-gauge
   scalar-harmonic regularization parameters to eliminate the leading-order
   even divergent powers of $\ell$ (e.g. $(2\ell+1)^2$ or $1$ depending on
   the case). These are the same as the tensor-harmonic parameters since
   the projection onto scalar harmonics only introduces a change in the
   regularization parameters at third-from-leading order and beyond.
\item For the tidal invariants, we are still left with the constant-in-$\ell$
   piece of the divergent metric perturbation. We identify this piece by
   taking an $\ell \to \infty$ limit of our PN expansions, and then subtract
   it to leave a convergent mode-sum.
\item Finally, we analytically sum over all $\ell$ modes (up to $\ell=\infty$)
   to obtain the regularized result.
\end{enumerate}

\section{Results}
\label{sec:results}
Our main result is the high-order post-Newtonian expansion of gauge invariant
quantities for a circular orbit extreme mass ratio binary system, ignoring
contributions from the spin of the constituent bodies. Similar post-Newtonian
expansions have been given to lower orders in other works
\cite{Detweiler:2008ft,Dolan:2013roa,Shah:2013uya,Dolan:2014pja,Bini:2013zaa,Bini:2013rfa,Bini:2014nfa,Bini:2014ica,Bini:2014zxa,Bini:2015bla,Bini:2015mza,Johnson-McDaniel:2015vva,Shah:2015nva}; we have verified that our
results are in perfect agreement with these.

Our post-Newtonian expressions for the invariants take a standard form in all
cases, and is given as a series in powers of $y$ and $\log y$. For example, the
spin-precession invariant is given by
\begin{align}
  \Delta \psi &= c_2 \, y^2 + c_3 \, y^3 + c_4 \, y^4 + (c_5 + c_5^{\rm ln} \log y) \, y^5 +
\nonumber \\
   &+ (c_6 + c_6^{\rm ln} \log y) \, y^6 + c_{6.5} \, y^{6.5} + (c_7 + c_7^{\rm ln} \log y) \, y^7
\nonumber \\
   &+ c_{7.5} \, y^{7.5} + (c_8 + c_8^{\rm ln} \log y + c_8^{{\rm ln}^2} \log^2 y)\,  y^8 \cdots,
\end{align}
where we use the notation that $c_{n}$ corresponds to the coefficient of $y^n$,
and $c_n^{{\rm ln}^k}$ is the coefficient of $y^n \log (y)^k$. The coefficients
are exact numerical quantities involving rational numbers, logs of integers,
$\pi$, Euler's constant ($\gamma$), and the Riemann zeta function ($\zeta$) with
integer argument (these various numerical quantities can be seen to come about
from the infinite sums over $\ell$ and $m$ spherical-harmonic modes). We give
the explicit values for many of the coefficients in the appendices, with
higher-order coefficients being available electronically \cite{online}.

In addition to gauge invariant quantities, our expressions may also be used to
compute a high-order post-Newtonian expansion of the linear-in-mass-ratio piece
of the effective one body radial interaction potential. As was shown in
Refs.~\cite{Barausse:2011dq,LeTiec:2011ab,LeTiec:2011dp}, this has a
straightforward expression in terms of the redshift invariant,
\begin{equation}
  a(y) = -(1-3y)^{3/2} \Delta U - \frac{y(1-4y)}{\sqrt{1-3y}},
\end{equation}
so substituting our post-Newtonian expansion for $\Delta U$ in to this
expression yields and expansion for $a(y)$.

\section{Discussion}
\label{sec:discussion}
Although we have given post-Newtonian expansions for all currently known
invariants for circular orbits in Schwarzschild spacetime, there remain further
quantities one could compute. For example, a forthcoming work \cite{Nolan:2015vpa} will
present numerical and post-Newtonian approximations for octupolar invariants
for circular orbits in Schwarzschild spacetime. Similarly our homogeneous solutions could
be used to calculate high order energy and angular momentum fluxes, as 
was done previously in for example \cite{2014PhRvD..90d4025S,2012PThPh.128..971F}. 
It is interesting to note that in an analysis of the form of the PN expansion of
the energy flux at infinity \cite{Johnson-McDaniel:2014lia}, Johnson-McDaniel identified the origin of many logarithmic and transcendental
terms in the flux, allowing significant simplification of the final expressions. 
It would be interesting to understand how these relate to our approach, 
and if any more efficiency could be added to our calculations by their observations.

From an astrophysical perspective a particularly compelling direction 
for future study is the extension to the Kerr background corresponding to a
rotating black hole. The derivation of the various gauge invariants for
circular orbits in Kerr spacetime has already been worked out in \cite{Dolan:2014pja}
and the MST method in general applies equally well to the Kerr case. The
outstanding issue is therefore the development of efficient post-Newtonian
expansion techniques for the Kerr case. Fortunately, we anticipate that many of
the tricks employed in this work will carry over to the Kerr case; we will
explore this in more detail in a future work.

\acknowledgments
We thank Bernard Whiting and Marc Casals for many fruitful discussions. We also
thank Donato Bini and Thibault Damour for helpful conversations, and Seth
Hopper and Patrick Nolan for some suggestions during the development of this
work. We are also grateful to Abhay Shah for providing his numerically-computed
expansion coefficients as a cross-check against the ones computed here.

This material is based upon work supported by the National Science Foundation
under Grant Number 1417132. B.W. was supported by the Irish Research Council,
which is funded under the National Development Plan for Ireland. C.K. is funded under the
Programme for Research in Third Level Institutions (PRTLI) Cycle 5 and
co-funded under the European Regional Development Fund.

\appendix

\begin{widetext}
\section{Gauge invariants}
\label{sec:gauge-invariants}
In this work we evaluate the redshift, spin-precession and tidal-tensor invariants by
writing them in terms of the components of $h_{ab}$ in Schwarzschild
coordinates. In particular, for our particular choice of off-worldline
extension the invariants are given by
\begin{equation}
  \Delta U = \frac{r_0}{r_0-3M} (h_{tt} + \Omega h_{t\phi} + \Omega^2 h_{\phi\phi}),
\end{equation}
\begin{align}
\label{eq:Delta-psi-Schwarzschild}
\Delta \psi = &
  \frac{1}{2 r_0 \Omega} \sqrt{\frac{r_0-3 M}{r_0}} \bigg[
      h_{tr,\phi}
    - h_{t\phi,r}
    + \Omega (h_{r\phi,\phi}-h_{\phi \phi,r} + f r_0 h_{rr})
    \bigg]
    \nonumber \\
    &
    +\frac{1}{2 M r_0 f} \sqrt{\frac{M}{r_0-3 M}} 
    \bigg[
      \Omega (
         M r_0^2 h_{tt}
        + r_0 f^2 h_{\phi \phi})
        + 2 M f h_{t\phi})
    \bigg],
\end{align}
\begin{align}
\label{eq:Delta-lambda1-Schwarzschild}
\Delta \lambda^E_1 = &
   \frac{\Omega^2 f (2 r_0 - 3 M)}{r_0 - 3 M} h_{rr}
   - \frac{\Omega^2(2 r_0^2-6 M r_0+3 M^2)}{f (r_0 - 3 M)^2} h_{tt}
   - \frac{6 M \Omega f h_{t\phi}}{r_0 (r_0 - 3 M)^2}
   - \frac{\Omega^2 (r_0^2-3 M r_0+3 M^2) h_{\phi\phi}}{r_0^2 (r_0 - 3 M)^2}
\nonumber \\
  &
 -\frac{r_0 - 2 M}{2 (r_0 - 3 M)} \Big[
     h_{tt,rr}
   + 2 \Omega h_{t\phi,rr}
   + \Omega^2 h_{\phi\phi,rr}
 \Big]
 -\frac{\Omega^2 h_{r\phi,\phi}
   + \Omega [h_{tr,\phi}+ h_{t\phi,r}]
   + h_{tt,r}}{r_0},
\end{align}
\begin{align}
\label{eq:Delta-lambda2-Schwarzschild}
\Delta \lambda^E_2 = &
  \frac{
    2 M [h_{t t} + 2\Omega h_{t \phi} + \Omega^2 h_{\phi \phi}]
    - [r_0 - 3 M] [
      h_{t t,\theta \theta}
      + 2\Omega h_{t \phi,\theta \theta}
      + \Omega^2 (h_{\phi \phi,\theta \theta} + 2 h_{\theta \theta})
    ]}{2r_0 (r_0 - 3 M)^2} ,
\end{align}
\begin{align}
\Delta \lambda^E_3 =
  \frac{\Omega^2}{f} h_{t t}
  -\Omega^2 f h_{r r}
  -\frac{\Omega^2}{r_0^2} h_{\phi \phi}
  +\frac{\Omega (h_{t \phi, r} - h_{t r, \phi})
    + \Omega^2(h_{\phi \phi, r} - h_{r \phi, \phi})}{r_0}
  -\frac{h_{t t, \phi \phi}
    + 2\Omega h_{t \phi, \phi \phi}
    + \Omega^2 h_{\phi \phi, \phi \phi}}{2 r_0^2 f},
\nonumber \\
\end{align}
\begin{align}
\label{eq:Delta-lambda3-Schwarzschild}
\Delta \lambda^B &=
 \frac{3\Omega^3 f^{1/2}}{(r_0-3 M)} h_{\theta\theta}
+\frac{\Omega^3 f^{1/2}(r_0-9 M)}{2(r_0-3 M)^2}h_{\phi\phi}
-\frac{\Omega^2 (r_0-M) h_{t\phi}}{\sqrt{f} (r_0-3 M)^2}
-\frac{\Omega M (5 r_0-9 M) h_{tt}}{2 \sqrt{f} r_0 (r_0-3 M)^2}
-\frac{\Omega f^{3/2}}{r_0 } h_{rr}
\nonumber \\
&
+\frac{\sqrt{f}}{2 r_0^2} \Big[
  \Omega [(h_{\theta\theta,r}
  -2 h_{r\theta,\theta})
  -h_{r\phi,\phi}]-h_{tr,\phi}
\Big]
+\frac{1}{2 \sqrt{f} r_0^3}\Big[(r_0-4 M) h_{t\phi,r} +\Omega (r_0-3 M) h_{\phi\phi,r}\Big]
-\frac{\Omega h_{tt,r}}{2 \sqrt{f}}
\nonumber \\
&
+\frac{\Omega}{2 \sqrt{f} r_0^2 (r_0-3 M)}\Big[
  f h_{\phi\phi,\theta\theta}+r_0^2 h_{tt,\theta\theta}\Big]
-\frac{1}{2 \sqrt{f} r_0^3} \Big[
  \Omega  h_{\theta\phi,\phi\theta}+h_{t\theta,\phi\theta}\Big]
+\frac{(r_0-M)}{2 \sqrt{f} r_0^3 (r_0-3 M)} h_{t\phi,\theta\theta}.
\end{align}
In the above we evaluate all of the metric perturbation components using
the expressions given in Sec.~\ref{sec:metric-reconstruction}, with
$(\theta,\phi) = (\tfrac{\pi}{2},0)$.

Note that other choices of off-worldline extension are possible. For example,
our expressions could have incorporated arbitrary extra dependence on $\theta$
and $\phi$, provided they smoothly approach the above expressions in the limit
to the worldline (see Refs.~\cite{Barack:2002bt,Dolan:2014pja,Wardell:2015kea}
for further discussion of this issue). However, we found the above to be the
most convenient to work with.

\section{Post-Newtonian expansion coefficients for \texorpdfstring{$\Delta U$}{Delta U}}
\label{sec:DeltaU}
In this Appendix, we give the analytic post-Newtonian expansion coefficients
for Detweiler's redshift invariant \cite{Detweiler:2008ft}, $\Delta U$, up to order $y^{15.5}$. We have also computed
higher order coefficients up to order $y^{21.5}$, but they are too
long to give here; instead, we make them available electronically
\cite{online}.

\begin{gather*}
c_{1} = -1, \quad
c_{2} = -2, \quad
c_{3} = -5, \quad
c_{4} = \tfrac{41}{32} \pi^2- \tfrac{121}{3},
\nonumber \\
c_{5} = - \tfrac{1157}{15}- \tfrac{128}{5} \gamma + \tfrac{677}{512} \pi^2- \tfrac{256}{5} \log(2), \quad
c_{5}^{{\rm ln}} = - \tfrac{64}{5},
\nonumber \\
c_{6} = \tfrac{1606877}{3150}+ \tfrac{1912}{105} \gamma - \tfrac{60343}{768} \pi^2+ \tfrac{7544}{105} \log(2)- \tfrac{243}{7} \log(3), \quad
c_{6}^{{\rm ln}} = \tfrac{956}{105}, \quad
c_{6.5} = - \tfrac{13696}{525} \pi,
\nonumber \\
c_{7} = \tfrac{17083661}{4050}+ \tfrac{102512}{567} \gamma - \tfrac{1246056911}{1769472} \pi^2+ \tfrac{2800873}{262144} \pi^4+ \tfrac{372784}{2835} \log(2)+ \tfrac{1215}{7} \log(3), \quad
c_{7}^{{\rm ln}} = \tfrac{51256}{567}, \quad
c_{7.5} = \tfrac{81077}{3675} \pi,
\end{gather*}


\section{Post-Newtonian expansion coefficients for \texorpdfstring{$\Delta \psi$}{Delta psi}}
\label{sec:DeltaPsi}
In this Appendix, we give the analytic post-Newtonian expansion coefficients
for the spin-precession invariant \cite{Dolan:2013roa}, $\Delta \psi$, up to order $y^{15.5}$. We have also computed
higher order coefficients up to order $y^{21.5}$, but they are too
long to give here; instead, we make them available electronically
\cite{online}.
\begin{gather*}
c_{2} = 1, \quad
c_{3} = -3, \quad
c_{4} = - \tfrac{15}{2}, \quad
c_{5} = - \frac{6277}{30} - 16 \gamma + \frac{20471}{1024} \pi^2 - \frac{496}{15} \log(2), \quad
c_{5}^{{\rm ln}} = -8,
\nonumber \\
c_{6} =  - \frac{87055}{28} - \frac{52}{5} \gamma + \frac{653629}{2048} \pi^2 - \frac{729}{14} \log(3) + \frac{3772}{105} \log(2), \quad
c_{6}^{{\rm ln}} = - \tfrac{26}{5}, \quad
c_{6.5} = - \tfrac{26536}{1575} \pi,
\nonumber \\
c_{7} = - \tfrac{149628163}{18900} + \tfrac{7628}{21} \gamma + \tfrac{297761947}{393216} \pi^2 - \tfrac{1407987}{524288} \pi^4 + \tfrac{4556}{21} \log(2) + \tfrac{12879}{35} \log(3),
\nonumber \\
c_{7}^{{\rm ln}} = \tfrac{3814}{21}, \quad
c_{7.5} = - \tfrac{113411}{22050} \pi,
\end{gather*}


\section{Post-Newtonian expansion coefficients for \texorpdfstring{$\Delta \lambda_1$}{Delta lambda 1}}
\label{sec:DeltaLambda1}
In this Appendix, we give the analytic post-Newtonian expansion coefficients
for the tidal invariant $\Delta \lambda_1$ (see \cite{Dolan:2014pja} for a definition) up to order $y^{15.5}$. We have also computed
higher order coefficients up to order $y^{21.5}$, but they are too
long to give here; instead, we make them available electronically \cite{online}.
\begin{gather*}
c_{3} = 2, \quad
c_{4} = 2, \quad
c_{5} = -\tfrac{19}{4}, \quad
c_{6} = \tfrac{227}{3}-\tfrac{593}{256} \pi^2, \quad
c_{7} = -\tfrac{71779}{4800}+\tfrac{768}{5} \gamma -\tfrac{719}{256} \pi^2+\tfrac{1536}{5} \log(2), \quad
c_{7}^{{\rm ln}} = \tfrac{384}{5},
\nonumber \\
c_{8} = \tfrac{35629703}{100800}-\tfrac{17152}{105} \gamma -\tfrac{1008787}{24576} \pi^2-\tfrac{5248}{7} \log(2)+\tfrac{2916}{7} \log(3), \quad
c_{8}^{{\rm ln}} = -\tfrac{8576}{105}, \quad
c_{8.5} = \tfrac{27392}{175} \pi,
\nonumber \\
c_{9} = -\tfrac{6746904013}{7257600}-\tfrac{5435624}{2835} \gamma +\tfrac{4692901483}{7077888} \pi^2+\tfrac{2193373}{1048576} \pi^4+\tfrac{877432}{2835} \log(2)-\tfrac{20898}{7} \log(3), \quad
c_{9}^{{\rm ln}} = -\tfrac{2717812}{2835},
\nonumber \\
c_{9.5} = -\tfrac{254116}{1225} \pi,
\end{gather*}
\begin{align}
c_{10} =& \tfrac{12288}{5} \zeta(3)-\tfrac{1964481413350639}{48898080000}+\tfrac{58241403128}{5457375} \gamma -\tfrac{219136}{175} \gamma^2+\tfrac{113134518813241}{19818086400} \pi^2-\tfrac{6653357405}{67108864} \pi^4-\tfrac{876544}{175} \log^2(2)
\nonumber \\
  & +\tfrac{6396680456}{5457375} \log(2)-\tfrac{876544}{175} \gamma \log(2)+\tfrac{6028101}{1120} \log(3)+\tfrac{9765625}{3168} \log(5),
\nonumber \\
c_{10}^{{\rm ln}} =& \tfrac{29120701564}{5457375}-\tfrac{219136}{175} \gamma -\tfrac{438272}{175} \log(2),
\nonumber \\
c_{10}^{{\rm ln}^2} =& -\tfrac{54784}{175},
\nonumber \\
c_{10.5} =& -\tfrac{5977039346}{3274425} \pi,
\nonumber \\
c_{11} =& \tfrac{84608}{15} \zeta(3)-\tfrac{125723025691048615381}{366148823040000}+\tfrac{169740444817}{4729725} \gamma +\tfrac{39488}{315} \gamma^2+\tfrac{385259238999617}{15854469120} \pi^2+\tfrac{40341026981537}{128849018880} \pi^4
\nonumber \\
  & +\tfrac{159488704}{11025} \log^2(2)-\tfrac{227448}{49} \log^2(3)+\tfrac{12895333428521}{165540375} \log(2)+\tfrac{36056192}{3675} \gamma \log(2)+\tfrac{7181878122}{111475} \log(3)
\nonumber \\
  & -\tfrac{454896}{49} \gamma \log(3)-\tfrac{454896}{49} \log(2) \log(3)-\tfrac{460937500}{11583} \log(5)+\tfrac{141087744}{17875} \log(6),
\nonumber \\
c_{11}^{{\rm ln}} =& \tfrac{170466930577}{9459450}+\tfrac{39488}{315} \gamma +\tfrac{18028096}{3675} \log(2)-\tfrac{227448}{49} \log(3),
\nonumber \\
c_{11}^{{\rm ln}^2} =& \tfrac{9872}{315},
\nonumber \\
c_{11.5} =& \tfrac{6426556598284309}{524431908000} \pi -\tfrac{46895104}{18375} \gamma \pi +\tfrac{438272}{525} \pi^3-\tfrac{93790208}{18375} \pi \log(2),
\nonumber \\
c_{11.5}^{{\rm ln}} =& -\tfrac{23447552}{18375} \pi,
\nonumber \\
c_{12} =& -\tfrac{98859104}{2835} \zeta(3)+\tfrac{53498906883627790209411067}{8373457434101760000}-\tfrac{53837173715700527}{442489422375} \gamma +\tfrac{169954100048}{9823275} \gamma^2+\tfrac{6912901078249843141}{73247647334400} \pi^2
\nonumber \\
  & -\tfrac{812988393706820631}{10995116277760} \pi^4-\tfrac{34346418841}{402653184} \pi^6-\tfrac{1124359929776}{9823275} \log^2(2)+\tfrac{1630044}{49} \log^2(3)+\tfrac{84315513161325592}{442489422375} \log(2)
\nonumber \\
  & -\tfrac{60665220064}{1964655} \gamma \log(2)-\tfrac{831114908742429}{1569568000} \log(3)+\tfrac{3260088}{49} \gamma \log(3)+\tfrac{3260088}{49} \log(2) \log(3)+\tfrac{11660541015625}{62270208} \log(5)
\nonumber \\
  & +\tfrac{678223072849}{34749000} \log(7),
\nonumber \\
c_{12}^{{\rm ln}} =& -\tfrac{53463358651678127}{884978844750}+\tfrac{169954100048}{9823275} \gamma -\tfrac{30332610032}{1964655} \log(2)+\tfrac{1630044}{49} \log(3),
\nonumber \\
c_{12}^{{\rm ln}^2} =& \tfrac{42488525012}{9823275},
\nonumber \\
c_{12.5} =& \tfrac{425532494729325719}{14913532383750} \pi +\tfrac{427856752}{385875} \gamma \pi +\tfrac{16089104}{11025} \pi^3+\tfrac{12561842768}{1157625} \pi \log(2)-\tfrac{2956824}{343} \pi \log(3),
\nonumber \\
c_{12.5}^{{\rm ln}} =& \tfrac{213928376}{385875} \pi,
\nonumber \\
c_{13} =& -\tfrac{2741094064}{5457375} \zeta(3)-\tfrac{7012352}{175} \gamma \zeta(3)-\tfrac{196608}{5} \zeta(5)-\tfrac{14024704}{175} \zeta(3) \log(2)+\tfrac{7136461453195733401088281131391999}{74095385081177017958400000}
\nonumber \\
  & -\tfrac{687851821144842957070451}{3011937000222150000} \gamma -\tfrac{3262516308400408}{49165491375} \gamma^2+\tfrac{375160832}{55125} \gamma^3+\tfrac{135867325661033176542256997}{234626863941550080000} \pi^2-\tfrac{93790208}{11025} \gamma \pi^2
\nonumber \\
  & -\tfrac{2398144603191276475096543}{2216615441596416000} \pi^4+\tfrac{267833739495301}{51539607552} \pi^6+\tfrac{3001286656}{55125} \log^3(2)+\tfrac{45668551753654888}{49165491375} \log^2(2)
\nonumber \\
  & +\tfrac{1500643328}{18375} \gamma \log^2(2)-\tfrac{1838920374351}{30830800} \log^2(3)-\tfrac{37744140625}{679536} \log^2(5)-\tfrac{11473878894172613656733927}{3011937000222150000} \log(2)
\nonumber \\
  & +\tfrac{14968695808903888}{49165491375} \gamma \log(2)+\tfrac{750321664}{18375} \gamma^2 \log(2)-\tfrac{187580416}{11025} \pi^2 \log(2)+\tfrac{54375724615556984499}{41971817888000} \log(3)
\nonumber \\
  & -\tfrac{1838920374351}{15415400} \gamma \log(3)-\tfrac{1838920374351}{15415400} \log(2) \log(3)+\tfrac{7781157422615234375}{11656173424896} \log(5)-\tfrac{37744140625}{339768} \gamma \log(5)
\nonumber \\
  & -\tfrac{37744140625}{339768} \log(2) \log(5)-\tfrac{3367765112736913}{9451728000} \log(7),
\nonumber \\
c_{13}^{{\rm ln}} =& -\tfrac{3506176}{175} \zeta(3)-\tfrac{672169095461400539470451}{6023874000444300000}-\tfrac{3262516308400408}{49165491375} \gamma +\tfrac{187580416}{18375} \gamma^2-\tfrac{46895104}{11025} \pi^2+\tfrac{750321664}{18375} \log^2(2)
\nonumber \\
  & +\tfrac{7484347904451944}{49165491375} \log(2)+\tfrac{750321664}{18375} \gamma \log(2)-\tfrac{1838920374351}{30830800} \log(3)-\tfrac{37744140625}{679536} \log(5),
\nonumber \\
c_{13}^{{\rm ln}^2} =& -\tfrac{815629077100102}{49165491375}+\tfrac{93790208}{18375} \gamma +\tfrac{187580416}{18375} \log(2),
\nonumber \\
c_{13}^{{\rm ln}^3} =& \tfrac{46895104}{55125},
\nonumber \\
c_{13.5} =& -\tfrac{425577146707570254105031}{3212732800236960000} \pi +\tfrac{928193260264}{28014525} \gamma \pi -\tfrac{30834184}{2695} \pi^3-\tfrac{80246336987576}{3094331625} \pi \log(2)+\tfrac{21190572}{343} \pi \log(3),
\nonumber \\
c_{13.5}^{{\rm ln}} =& \tfrac{464096630132}{28014525} \pi,
\nonumber \\
c_{14} =& \tfrac{100736677024996}{70945875} \zeta(3)-\tfrac{285620224}{1575} \gamma \zeta(3)-\tfrac{5848064}{15} \zeta(5)-\tfrac{104290304}{3675} \zeta(3) \log(2)-\tfrac{16376256}{49} \zeta(3) \log(3)
\nonumber \\
  & +\tfrac{14782927091165826490177755897290448751}{17157712607860053220992000000}+\tfrac{667577042987116745835126713}{247982813018290350000} \gamma -\tfrac{467801606965920698}{1065252313125} \gamma^2+\tfrac{4226696704}{496125} \gamma^3
\nonumber \\
  & -\tfrac{475545050053701762012976804979}{878442978597163499520000} \pi^2-\tfrac{1056674176}{99225} \gamma \pi^2-\tfrac{10521866887485742964118830609}{1276770494359535616000} \pi^4-\tfrac{5695388196662513281}{103903848824832} \pi^6
\nonumber \\
  & -\tfrac{28666494464}{165375} \log^3(2)+\tfrac{11827296}{343} \log^3(3)-\tfrac{9125991752151527542}{2033663506875} \log^2(2)-\tfrac{48218332672}{231525} \gamma \log^2(2)
\nonumber \\
  & +\tfrac{35481888}{343} \log^2(2) \log(3)-\tfrac{5598682936438608}{6137256125} \log^2(3)+\tfrac{35481888}{343} \gamma \log^2(3)+\tfrac{35481888}{343} \log(2) \log^2(3)
\nonumber \\
  & +\tfrac{3563046875000}{4969107} \log^2(5)+\tfrac{78947326245948789380775445087}{5207639073384097350000} \log(2)-\tfrac{63277612176600694372}{22370298575625} \gamma \log(2)-\tfrac{6737700352}{128625} \gamma^2 \log(2)
\nonumber \\
  & +\tfrac{1684425088}{77175} \pi^2 \log(2)+\tfrac{596058807449014872782742513}{81277585903754240000} \log(3)-\tfrac{11197365872877216}{6137256125} \gamma \log(3)+\tfrac{35481888}{343} \gamma^2 \log(3)
\nonumber \\
  & -\tfrac{14784120}{343} \pi^2 \log(3)-\tfrac{13290270496988832}{6137256125} \log(2) \log(3)+\tfrac{70963776}{343} \gamma \log(2) \log(3)-\tfrac{1298204231280352501953125}{103646694094175232} \log(5)
\nonumber \\
  & +\tfrac{7126093750000}{4969107} \gamma \log(5)+\tfrac{7126093750000}{4969107} \log(2) \log(5)+\tfrac{213966604377170305829}{81497954304000} \log(7),
\nonumber \\
c_{14}^{{\rm ln}} =& -\tfrac{142810112}{1575} \zeta(3)+\tfrac{674213554413595385800585913}{495965626036580700000}-\tfrac{467801606965920698}{1065252313125} \gamma +\tfrac{2113348352}{165375} \gamma^2-\tfrac{528337088}{99225} \pi^2
\nonumber \\
  & -\tfrac{24109166336}{231525} \log^2(2)+\tfrac{17740944}{343} \log^2(3)-\tfrac{31666818304007022386}{22370298575625} \log(2)-\tfrac{6737700352}{128625} \gamma \log(2)
\nonumber \\
  & -\tfrac{5598682936438608}{6137256125} \log(3)+\tfrac{35481888}{343} \gamma \log(3)+\tfrac{35481888}{343} \log(2) \log(3)+\tfrac{3563046875000}{4969107} \log(5),
\nonumber \\
c_{14}^{{\rm ln}^2} =& -\tfrac{235234718516611549}{2130504626250}+\tfrac{1056674176}{165375} \gamma -\tfrac{1684425088}{128625} \log(2)+\tfrac{8870472}{343} \log(3),
\nonumber \\
c_{14}^{{\rm ln}^3} =& \tfrac{528337088}{496125},
\nonumber \\
c_{14.5} =& -\tfrac{750321664}{18375} \pi \zeta(3)-\tfrac{9637041144752939728799097230831}{233718841613478289068000000} \pi -\tfrac{4846441673897531348153}{29528794119825000} \gamma \pi +\tfrac{40142209024}{1929375} \gamma^2 \pi 
\nonumber \\
  & +\tfrac{9374390487730897}{1966619655000} \pi^3-\tfrac{750321664}{55125} \gamma \pi^3-\tfrac{7012352}{7875} \pi^5+\tfrac{160568836096}{1929375} \pi \log^2(2)+\tfrac{16831992759911954495957}{88586382359475000} \pi \log(2)
\nonumber \\
  & +\tfrac{160568836096}{1929375} \gamma \pi \log(2)-\tfrac{1500643328}{55125} \pi^3 \log(2)-\tfrac{17024692325983821}{154308154000} \pi \log(3)-\tfrac{29176220703125}{291520944} \pi \log(5),
\nonumber \\
c_{14.5}^{{\rm ln}} =& -\tfrac{4846441673897531348153}{59057588239650000} \pi +\tfrac{40142209024}{1929375} \gamma \pi -\tfrac{375160832}{55125} \pi^3+\tfrac{80284418048}{1929375} \pi \log(2),
\nonumber \\
c_{14.5}^{{\rm ln}^2} =& \tfrac{10035552256}{1929375} \pi,
\nonumber \\
c_{15} =& \tfrac{821662588800605947}{189638323875} \zeta(3)+\tfrac{384897840256}{1403325} \gamma \zeta(3)-\tfrac{341588608}{405} \zeta(5)-\tfrac{39225871528064}{9823275} \zeta(3) \log(2)+\tfrac{117363168}{49} \zeta(3) \log(3)
\nonumber \\
  & -\tfrac{18422245917662946965713688497238514272871305051}{1131032022735707096154168066048000000}+\tfrac{195356603278656058756665560679851}{22537174012728263588700000} \gamma +\tfrac{27824356547133741211189}{96812260721426250} \gamma^2
\nonumber \\
  & -\tfrac{1307323995299008}{14587563375} \gamma^3+\tfrac{1271505627196870617621895639332984239}{2419231963056588277678080000} \pi^2+\tfrac{326830998824752}{2917512675} \gamma \pi^2
\nonumber \\
  & +\tfrac{632727847482480999735886759645493}{19415082845428842391142400} \pi^4+\tfrac{96600643595949949145287}{11821949021847552} \pi^6+\tfrac{47849384934497471}{9895604649984} \pi^8
\nonumber \\
  & +\tfrac{83625371578984384}{20422588725} \log^3(2)-\tfrac{84762288}{343} \log^3(3)-\tfrac{12693821815837334753855293}{677685825049983750} \log^2(2)+\tfrac{121913594374637248}{34037647875} \gamma \log^2(2)
\nonumber \\
  & -\tfrac{254286864}{343} \log^2(2) \log(3)+\tfrac{2903575106583115821}{395108560000} \log^2(3)-\tfrac{254286864}{343} \gamma \log^2(3)-\tfrac{254286864}{343} \log(2) \log^2(3)
\nonumber \\
  & -\tfrac{5359327017958984375}{1589478194304} \log^2(5)-\tfrac{14180966230199741}{28798233750} \log^2(7)+\tfrac{23991851361324447286645878839586053}{631040872356391380483600000} \log(2)
\nonumber \\
  & -\tfrac{2528027615481223322081309}{338842912524991875} \gamma \log(2)+\tfrac{23495190959122112}{34037647875} \gamma^2 \log(2)-\tfrac{5873797739780528}{20422588725} \pi^2 \log(2)
\nonumber \\
  & -\tfrac{45919835559509188340293117947}{531430369370700800000} \log(3)+\tfrac{2903575106583115821}{197554280000} \gamma \log(3)-\tfrac{254286864}{343} \gamma^2 \log(3)+\tfrac{105952860}{343} \pi^2 \log(3)
\nonumber \\
  & +\tfrac{51416702942634990153}{2568205640000} \log(2) \log(3)-\tfrac{508573728}{343} \gamma \log(2) \log(3)+\tfrac{75097296400459024163587890625}{1332067312498340081664} \log(5)
\nonumber \\
  & -\tfrac{5359327017958984375}{794739097152} \gamma \log(5)-\tfrac{5359327017958984375}{794739097152} \log(2) \log(5)+\tfrac{439852610132865166823002613}{294210467465840640000} \log(7)
\nonumber \\
  & -\tfrac{14180966230199741}{14399116875} \gamma \log(7)-\tfrac{14180966230199741}{14399116875} \log(2) \log(7),
\nonumber \\
c_{15}^{{\rm ln}} =& \tfrac{192448920128}{1403325} \zeta(3)+\tfrac{198076227935385570494604925311851}{45074348025456527177400000}+\tfrac{27316781481642589953589}{96812260721426250} \gamma -\tfrac{653661997649504}{4862521125} \gamma^2
\nonumber \\
  & +\tfrac{163415499412376}{2917512675} \pi^2+\tfrac{60956797187318624}{34037647875} \log^2(2)-\tfrac{127143432}{343} \log^2(3)-\tfrac{2539376676711189532231709}{677685825049983750} \log(2)
\nonumber \\
  & +\tfrac{23495190959122112}{34037647875} \gamma \log(2)+\tfrac{2903575106583115821}{395108560000} \log(3)-\tfrac{254286864}{343} \gamma \log(3)-\tfrac{254286864}{343} \log(2) \log(3)
\nonumber \\
  & -\tfrac{5359327017958984375}{1589478194304} \log(5)-\tfrac{14180966230199741}{28798233750} \log(7),
\nonumber \\
c_{15}^{{\rm ln}^2} =& \tfrac{25596917755268555111989}{387249042885705000}-\tfrac{326830998824752}{4862521125} \gamma +\tfrac{5873797739780528}{34037647875} \log(2)-\tfrac{63571716}{343} \log(3),
\nonumber \\
c_{15}^{{\rm ln}^3} =& -\tfrac{163415499412376}{14587563375},
\nonumber \\
c_{15.5} =& -\tfrac{19873715648}{128625} \pi \zeta(3)+\tfrac{22009514064102963901282098495898331237}{8620797379193684518369536000000} \pi -\tfrac{83840515418574757567694}{111963344371003125} \gamma \pi +\tfrac{674951603104}{40516875} \gamma^2 \pi 
\nonumber \\
  & +\tfrac{3153187507668059882}{7456766191875} \pi^3-\tfrac{19873715648}{385875} \gamma \pi^3-\tfrac{1280364352}{165375} \pi^5-\tfrac{8979249667168}{40516875} \pi \log^2(2)+\tfrac{230632272}{2401} \pi \log^2(3)
\nonumber \\
  & -\tfrac{343304287588063111961138}{143952871334146875} \pi \log(2)-\tfrac{435962654912}{3472875} \gamma \pi \log(2)+\tfrac{45397696}{99225} \pi^3 \log(2)-\tfrac{2122714909893642552}{1228678676225} \pi \log(3)
\nonumber \\
  & +\tfrac{461264544}{2401} \gamma \pi \log(3)-\tfrac{35481888}{343} \pi^3 \log(3)+\tfrac{461264544}{2401} \pi \log(2) \log(3)+\tfrac{2754235234375000}{2131746903} \pi \log(5),
\nonumber \\
c_{15.5}^{{\rm ln}} =& -\tfrac{41920257709287378783847}{111963344371003125} \pi +\tfrac{674951603104}{40516875} \gamma \pi -\tfrac{9936857824}{385875} \pi^3-\tfrac{217981327456}{3472875} \pi \log(2)+\tfrac{230632272}{2401} \pi \log(3),
\nonumber \\
c_{15.5}^{{\rm ln}^2} =& \tfrac{168737900776}{40516875} \pi.
\end{align}

\section{Post-Newtonian expansion coefficients for \texorpdfstring{$\Delta \lambda_2$}{Delta lambda 2}}
\label{sec:DeltaLambda2}
In this Appendix, we give the analytic post-Newtonian expansion coefficients
for the tidal invariant $\Delta \lambda_2$ (see \cite{Dolan:2014pja} for a definition) up to order $y^{15.5}$. We have also computed
higher order coefficients up to order $y^{21.5}$, but they are too
long to give here; instead, we make them available electronically \cite{online}.
\begin{gather*}
c_{3} = -1, \quad
c_{4} = -\tfrac{3}{2}, \quad
c_{5} = -\tfrac{23}{8}, \quad
c_{6} = \tfrac{1249}{1024} \pi^2-\tfrac{2593}{48}, \quad
c_{7} = -\tfrac{362051}{3200}-\tfrac{256}{5} \gamma +\tfrac{1737}{1024} \pi^2-\tfrac{512}{5} \log(2),
\nonumber \\
c_{7}^{{\rm ln}} = -\tfrac{128}{5}, \quad
c_{8} = \tfrac{917879}{1280}+\tfrac{176}{7} \gamma -\tfrac{7637151}{65536} \pi^2+\tfrac{16592}{105} \log(2)-\tfrac{729}{7} \log(3), \quad
c_{8}^{{\rm ln}} = \tfrac{88}{7}, \quad
c_{8.5} = -\tfrac{27392}{525} \pi,
\nonumber \\
c_{9} = \tfrac{35725395527}{2903040}+\tfrac{1193824}{2835} \gamma -\tfrac{24327985735}{14155776} \pi^2+\tfrac{29225393}{2097152} \pi^4+\tfrac{2368}{405} \log(2)+\tfrac{1215}{2} \log(3), \quad
c_{9}^{{\rm ln}} = \tfrac{596912}{2835}, \quad
c_{9.5} = \tfrac{58087}{1575} \pi,
\end{gather*}
\begin{align}
c_{10} =& -\tfrac{4096}{5} \zeta(3)+\tfrac{14160351233504683}{130394880000}-\tfrac{6572909192}{1819125} \gamma +\tfrac{219136}{525} \gamma^2-\tfrac{282633370752961}{26424115200} \pi^2-\tfrac{56292563931}{536870912} \pi^4+\tfrac{876544}{525} \log^2(2)
\nonumber \\
  & -\tfrac{7034284984}{1819125} \log(2)+\tfrac{876544}{525} \gamma \log(2)-\tfrac{2838483}{3520} \log(3)-\tfrac{9765625}{19008} \log(5),
\nonumber \\
c_{10}^{{\rm ln}} =& -\tfrac{3286454596}{1819125}+\tfrac{219136}{525} \gamma +\tfrac{438272}{525} \log(2),
\nonumber \\
c_{10}^{{\rm ln}^2} =& \tfrac{54784}{525},
\nonumber \\
c_{10.5} =& \tfrac{2672297839}{6548850} \pi,
\nonumber \\
c_{11} =& -\tfrac{57312}{35} \zeta(3)-\tfrac{4696484502278412099877}{5126083522560000}-\tfrac{4893706340371}{496621125} \gamma +\tfrac{931408}{11025} \gamma^2-\tfrac{47818652586050411}{1109812838400} \pi^2+\tfrac{1179941921594993}{85899345920} \pi^4
\nonumber \\
  & -\tfrac{34897328}{11025} \log^2(2)+\tfrac{56862}{49} \log^2(3)-\tfrac{361932668021}{23648625} \log(2)-\tfrac{3146144}{1575} \gamma \log(2)-\tfrac{13359835171323}{784784000} \log(3)
\nonumber \\
  & +\tfrac{113724}{49} \gamma \log(3)+\tfrac{113724}{49} \log(2) \log(3)+\tfrac{8826171875}{1482624} \log(5),
\nonumber \\
c_{11}^{{\rm ln}} =& -\tfrac{4919133341971}{993242250}+\tfrac{931408}{11025} \gamma -\tfrac{1573072}{1575} \log(2)+\tfrac{56862}{49} \log(3),
\nonumber \\
c_{11}^{{\rm ln}^2} =& \tfrac{232852}{11025},
\nonumber \\
c_{11.5} =& -\tfrac{1425524472919397}{349621272000} \pi +\tfrac{46895104}{55125} \gamma \pi -\tfrac{438272}{1575} \pi^3+\tfrac{93790208}{55125} \pi \log(2),
\nonumber \\
c_{11.5}^{{\rm ln}} =& \tfrac{23447552}{55125} \pi,
\nonumber \\
c_{12} =& \tfrac{4433552}{567} \zeta(3)-\tfrac{100903879415147484641886367}{3044893612400640000}+\tfrac{341935703082793}{12872419560} \gamma -\tfrac{3437020472}{893025} \gamma^2-\tfrac{18276670153689478799}{106542032486400} \pi^2
\nonumber \\
  & +\tfrac{145187213724599353117}{395824185999360} \pi^4-\tfrac{3328032926399}{3221225472} \pi^6+\tfrac{42570053816}{1964655} \log^2(2)-\tfrac{47385}{7} \log^2(3)-\tfrac{1564401033343}{31755240} \log(2)
\nonumber \\
  & +\tfrac{6794600272}{1403325} \gamma \log(2)+\tfrac{471959547237}{4928000} \log(3)-\tfrac{94770}{7} \gamma \log(3)-\tfrac{94770}{7} \log(2) \log(3)-\tfrac{341119140625}{13837824} \log(5)
\nonumber \\
  & -\tfrac{678223072849}{277992000} \log(7),
\nonumber \\
c_{12}^{{\rm ln}} =& \tfrac{336663160031017}{25744839120}-\tfrac{3437020472}{893025} \gamma +\tfrac{3397300136}{1403325} \log(2)-\tfrac{47385}{7} \log(3),
\nonumber \\
c_{12}^{{\rm ln}^2} =& -\tfrac{859255118}{893025},
\nonumber \\
c_{12.5} =& -\tfrac{2608988253558091529}{318155357520000} \pi -\tfrac{1940124}{42875} \gamma \pi -\tfrac{14576396}{33075} \pi^3-\tfrac{2608166516}{1157625} \pi \log(2)+\tfrac{739206}{343} \pi \log(3),
\nonumber \\
c_{12.5}^{{\rm ln}} =& -\tfrac{970062}{42875} \pi,
\nonumber \\
c_{13} =& -\tfrac{49988601004}{1819125} \zeta(3)+\tfrac{7012352}{525} \gamma \zeta(3)+\tfrac{65536}{5} \zeta(5)+\tfrac{14024704}{525} \zeta(3) \log(2)-\tfrac{19849728645586798963670301395715143}{49396923387451345305600000}
\nonumber \\
  & -\tfrac{1625662377940787050427}{55776611115225000} \gamma +\tfrac{144072014504954}{5462832375} \gamma^2-\tfrac{375160832}{165375} \gamma^3+\tfrac{316372978729180099591649107}{469253727883100160000} \pi^2+\tfrac{93790208}{33075} \gamma \pi^2
\nonumber \\
  & +\tfrac{49196714267884383672303691}{13299692649578496000} \pi^4+\tfrac{1224709581058837}{34359738368} \pi^6-\tfrac{3001286656}{165375} \log^3(2)-\tfrac{1575142519778962}{16388497125} \log^2(2)
\nonumber \\
  & -\tfrac{1500643328}{55125} \gamma \log^2(2)+\tfrac{77840476011}{8808800} \log^2(3)+\tfrac{37744140625}{4077216} \log^2(5)+\tfrac{267827958591244578422639}{501989500037025000} \log(2)
\nonumber \\
  & +\tfrac{169787211215068}{16388497125} \gamma \log(2)-\tfrac{750321664}{55125} \gamma^2 \log(2)+\tfrac{187580416}{33075} \pi^2 \log(2)-\tfrac{27023001987617566479}{167887271552000} \log(3)
\nonumber \\
  & +\tfrac{77840476011}{4404400} \gamma \log(3)+\tfrac{77840476011}{4404400} \log(2) \log(3)-\tfrac{18234546953814453125}{139874081098752} \log(5)+\tfrac{37744140625}{2038608} \gamma \log(5)
\nonumber \\
  & +\tfrac{37744140625}{2038608} \log(2) \log(5)+\tfrac{43503165672743}{1050192000} \log(7),
\nonumber \\
c_{13}^{{\rm ln}} =& \tfrac{3506176}{525} \zeta(3)-\tfrac{1776762811069392010427}{111553222230450000}+\tfrac{144072014504954}{5462832375} \gamma -\tfrac{187580416}{55125} \gamma^2+\tfrac{46895104}{33075} \pi^2-\tfrac{750321664}{55125} \log^2(2)
\nonumber \\
  & +\tfrac{84893605607534}{16388497125} \log(2)-\tfrac{750321664}{55125} \gamma \log(2)+\tfrac{77840476011}{8808800} \log(3)+\tfrac{37744140625}{4077216} \log(5),
\nonumber \\
c_{13}^{{\rm ln}^2} =& \tfrac{72036007252477}{10925664750}-\tfrac{93790208}{55125} \gamma -\tfrac{187580416}{55125} \log(2),
\nonumber \\
c_{13}^{{\rm ln}^3} =& -\tfrac{46895104}{165375},
\nonumber \\
c_{13.5} =& \tfrac{336629234198810065383583}{11565838080853056000} \pi -\tfrac{84454061934086}{11345882625} \gamma \pi +\tfrac{25421917126}{9823275} \pi^3+\tfrac{731692656242}{194500845} \pi \log(2)-\tfrac{616005}{49} \pi \log(3),
\nonumber \\
c_{13.5}^{{\rm ln}} =& -\tfrac{42227030967043}{11345882625} \pi,
\nonumber \\
c_{14} =& -\tfrac{198044268661346}{496621125} \zeta(3)+\tfrac{539634176}{11025} \gamma \zeta(3)+\tfrac{707072}{7} \zeta(5)+\tfrac{157460992}{11025} \zeta(3) \log(2)+\tfrac{4094064}{49} \zeta(3) \log(3)
\nonumber \\
  & +\tfrac{1587976079797604814918786653278459573361}{854072805369033760333824000000}-\tfrac{8723412958458566087930441177}{10415278146768194700000} \gamma +\tfrac{2661520081340964401}{22370298575625} \gamma^2-\tfrac{9525895936}{3472875} \gamma^3
\nonumber \\
  & -\tfrac{334704305890999854905574053879033}{3513771914388653998080000} \pi^2+\tfrac{2381473984}{694575} \gamma \pi^2+\tfrac{240872323127917332868816475237}{47666098456089329664000} \pi^4
\nonumber \\
  & -\tfrac{203477240692779353935}{138538465099776} \pi^6+\tfrac{44532653824}{1157625} \log^3(2)-\tfrac{2956824}{343} \log^3(3)+\tfrac{4868496539064319883}{7456766191875} \log^2(2)
\nonumber \\
  & +\tfrac{51812900096}{1157625} \gamma \log^2(2)-\tfrac{8870472}{343} \log^2(2) \log(3)+\tfrac{81604845977535927}{392784392000} \log^2(3)
\nonumber \\
  & -\tfrac{8870472}{343} \gamma \log^2(3)-\tfrac{8870472}{343} \log(2) \log^2(3)-\tfrac{34113154296875}{318022848} \log^2(5)-\tfrac{13970634868229722249737557}{5063334052877100000} \log(2)
\nonumber \\
  & +\tfrac{987397297378547014}{2033663506875} \gamma \log(2)+\tfrac{3640123136}{385875} \gamma^2 \log(2)-\tfrac{910030784}{231525} \pi^2 \log(2)-\tfrac{15913308806292853013647257}{9562068929853440000} \log(3)
\nonumber \\
  & +\tfrac{81604845977535927}{196392196000} \gamma \log(3)-\tfrac{8870472}{343} \gamma^2 \log(3)+\tfrac{3696030}{343} \pi^2 \log(3)+\tfrac{91172409973474743}{196392196000} \log(2) \log(3)
\nonumber \\
  & -\tfrac{17740944}{343} \gamma \log(2) \log(3)+\tfrac{774882301042051373046875}{414586776376700928} \log(5)-\tfrac{34113154296875}{159011424} \gamma \log(5)
\nonumber \\
  & -\tfrac{34113154296875}{159011424} \log(2) \log(5)-\tfrac{504631202715118986059}{1792954994688000} \log(7),
\nonumber \\
c_{14}^{{\rm ln}} =& \tfrac{269817088}{11025} \zeta(3)-\tfrac{8873669383459120101460629977}{20830556293536389400000}+\tfrac{2661520081340964401}{22370298575625} \gamma -\tfrac{4762947968}{1157625} \gamma^2+\tfrac{1190736992}{694575} \pi^2
\nonumber \\
  & +\tfrac{25906450048}{1157625} \log^2(2)-\tfrac{4435236}{343} \log^2(3)+\tfrac{494547503710687907}{2033663506875} \log(2)+\tfrac{3640123136}{385875} \gamma \log(2)
\nonumber \\
  & +\tfrac{81604845977535927}{392784392000} \log(3)-\tfrac{8870472}{343} \gamma \log(3)-\tfrac{8870472}{343} \log(2) \log(3)-\tfrac{34113154296875}{318022848} \log(5),
\nonumber \\
c_{14}^{{\rm ln}^2} =& \tfrac{2680194891812081201}{89481194302500}-\tfrac{2381473984}{1157625} \gamma +\tfrac{910030784}{385875} \log(2)-\tfrac{2217618}{343} \log(3),
\nonumber \\
c_{14}^{{\rm ln}^3} =& -\tfrac{1190736992}{3472875},
\nonumber \\
c_{14.5} =& \tfrac{750321664}{55125} \pi \zeta(3)-\tfrac{4247632379724880440936861021251}{54992668614936068016000000} \pi +\tfrac{3627715358766651040027}{59057588239650000} \gamma \pi -\tfrac{40142209024}{5788125} \gamma^2 \pi 
\nonumber \\
  & -\tfrac{12540743342866241}{1311079770000} \pi^3+\tfrac{750321664}{165375} \gamma \pi^3+\tfrac{7012352}{23625} \pi^5-\tfrac{160568836096}{5788125} \pi \log^2(2)+\tfrac{239685078235585923331}{6561954248850000} \pi \log(2)
\nonumber \\
  & -\tfrac{160568836096}{5788125} \gamma \pi \log(2)+\tfrac{1500643328}{165375} \pi^3 \log(2)+\tfrac{28590177065193}{1763521760} \pi \log(3)+\tfrac{29176220703125}{1749125664} \pi \log(5),
\nonumber \\
c_{14.5}^{{\rm ln}} =& \tfrac{3627715358766651040027}{118115176479300000} \pi -\tfrac{40142209024}{5788125} \gamma \pi +\tfrac{375160832}{165375} \pi^3-\tfrac{80284418048}{5788125} \pi \log(2),
\nonumber \\
c_{14.5}^{{\rm ln}^2} =& -\tfrac{10035552256}{5788125} \pi,
\nonumber \\
c_{15} =& -\tfrac{3988878460574807099}{5309873068500} \zeta(3)-\tfrac{141597386368}{1964655} \gamma \zeta(3)+\tfrac{418238336}{2835} \zeta(5)+\tfrac{7588521025408}{9823275} \zeta(3) \log(2)-\tfrac{3411720}{7} \zeta(3) \log(3)
\nonumber \\
  & +\tfrac{2414892567708773463979964883629243110584543483233}{15834448318299899346158352924672000000}-\tfrac{495857514022424850982475047313807}{315520436178195690241800000} \gamma
\nonumber \\
  &  -\tfrac{253435635132530765761019}{2710743300199935000} \gamma^2+\tfrac{416584131877696}{20422588725} \gamma^3-\tfrac{725545905960663924782740410707463647}{172802283075470591262720000} \pi^2-\tfrac{104146032969424}{4084517745} \gamma \pi^2
\nonumber \\
  & -\tfrac{695532337186146671394101506426105447}{1359055799180018967379968000} \pi^4-\tfrac{1972817957393892394411583}{30399297484750848} \pi^6+\tfrac{2451925837282700461}{19791209299968} \pi^8
\nonumber \\
  & -\tfrac{11744910335224384}{14587563375} \log^3(2)+\tfrac{2464020}{49} \log^3(3)+\tfrac{12426993372195597130915781}{2710743300199935000} \log^2(2)
\nonumber \\
  & -\tfrac{23633623737624512}{34037647875} \gamma \log^2(2)+\tfrac{7392060}{49} \log^2(2) \log(3)-\tfrac{24210899307284295303}{19078099040000} \log^2(3)+\tfrac{7392060}{49} \gamma \log^2(3)
\nonumber \\
  & +\tfrac{7392060}{49} \log(2) \log^2(3)+\tfrac{1409258865224609375}{3178956388608} \log^2(5)+\tfrac{14180966230199741}{230385870000} \log^2(7)
\nonumber \\
  & -\tfrac{5847778524718005712851407467454953}{631040872356391380483600000} \log(2)+\tfrac{2631218093161754315652421}{1355371650099967500} \gamma \log(2)-\tfrac{4280895102076864}{34037647875} \gamma^2 \log(2)
\nonumber \\
  & +\tfrac{1070223775519216}{20422588725} \pi^2 \log(2)+\tfrac{51234996291683954427554286951}{3947768458182348800000} \log(3)-\tfrac{24210899307284295303}{9539049520000} \gamma \log(3)
\nonumber \\
  & +\tfrac{7392060}{49} \gamma^2 \log(3)-\tfrac{3080025}{49} \pi^2 \log(3)-\tfrac{30860681711808573063}{9539049520000} \log(2) \log(3)+\tfrac{14784120}{49} \gamma \log(2) \log(3)
\nonumber \\
  & -\tfrac{19411060079856302297634765625}{2664134624996680163328} \log(5)+\tfrac{1409258865224609375}{1589478194304} \gamma \log(5)+\tfrac{1409258865224609375}{1589478194304} \log(2) \log(5)
\nonumber \\
  & -\tfrac{529932652475491438929787129}{1176841869863362560000} \log(7)+\tfrac{14180966230199741}{115192935000} \gamma \log(7)+\tfrac{14180966230199741}{115192935000} \log(2) \log(7),
\nonumber \\
c_{15}^{{\rm ln}} =& -\tfrac{70798693184}{1964655} \zeta(3)-\tfrac{517442198628498393725473546203407}{631040872356391380483600000}-\tfrac{248698267854613354023419}{2710743300199935000} \gamma +\tfrac{208292065938848}{6807529575} \gamma^2
\nonumber \\
  & -\tfrac{52073016484712}{4084517745} \pi^2-\tfrac{11816811868812256}{34037647875} \log^2(2)+\tfrac{3696030}{49} \log^2(3)+\tfrac{2647623077895636290282821}{2710743300199935000} \log(2)
\nonumber \\
  & -\tfrac{4280895102076864}{34037647875} \gamma \log(2)-\tfrac{24210899307284295303}{19078099040000} \log(3)+\tfrac{7392060}{49} \gamma \log(3)+\tfrac{7392060}{49} \log(2) \log(3)
\nonumber \\
  & +\tfrac{1409258865224609375}{3178956388608} \log(5)+\tfrac{14180966230199741}{230385870000} \log(7),
\nonumber \\
c_{15}^{{\rm ln}^2} =& -\tfrac{230100400220601639975419}{10842973200799740000}+\tfrac{104146032969424}{6807529575} \gamma -\tfrac{1070223775519216}{34037647875} \log(2)+\tfrac{1848015}{49} \log(3),
\nonumber \\
c_{15}^{{\rm ln}^3} =& \tfrac{52073016484712}{20422588725},
\nonumber \\
c_{15.5} =& \tfrac{2331205552}{55125} \pi \zeta(3)-\tfrac{17814262485036076084068117543498025079}{22167764689355188761521664000000} \pi +\tfrac{959470947789066888806999}{4606491882692700000} \gamma \pi -\tfrac{104717532296}{17364375} \gamma^2 \pi 
\nonumber \\
  & -\tfrac{451479164965426159}{3787563780000} \pi^3+\tfrac{2331205552}{165375} \gamma \pi^3+\tfrac{142842128}{70875} \pi^5+\tfrac{5830228449352}{121550625} \pi \log^2(2)-\tfrac{57658068}{2401} \pi \log^2(3)
\nonumber \\
  & +\tfrac{4347734456755030650576587}{10748481059616300000} \pi \log(2)+\tfrac{323272814608}{13505625} \gamma \pi \log(2)+\tfrac{8062785904}{3472875} \pi^3 \log(2)+\tfrac{777809311580363643621}{1965885881960000} \pi \log(3)
\nonumber \\
  & -\tfrac{115316136}{2401} \gamma \pi \log(3)+\tfrac{8870472}{343} \pi^3 \log(3)-\tfrac{115316136}{2401} \pi \log(2) \log(3)-\tfrac{26369468271484375}{136431801792} \pi \log(5),
\nonumber \\
c_{15.5}^{{\rm ln}} =& \tfrac{959470947789066888806999}{9212983765385400000} \pi -\tfrac{104717532296}{17364375} \gamma \pi +\tfrac{1165602776}{165375} \pi^3+\tfrac{161636407304}{13505625} \pi \log(2)-\tfrac{57658068}{2401} \pi \log(3),
\nonumber \\
c_{15.5}^{{\rm ln}^2} =& -\tfrac{26179383074}{17364375} \pi.
\end{align}

\section{Post-Newtonian expansion coefficients for \texorpdfstring{$\Delta \lambda_3$}{Delta lambda 3}}
\label{sec:DeltaLambda3}
In this Appendix, we give the analytic post-Newtonian expansion coefficients
for the tidal invariant $\Delta \lambda_3$ (see \cite{Dolan:2014pja} for a definition) up to order $y^{15.5}$. We have also computed
higher order coefficients up to order $y^{21.5}$, but they are too
long to give here; instead, we make them available electronically \cite{online}.
\begin{gather*}
c_{3} = -1, \quad
c_{4} = -\tfrac{1}{2}, \quad
c_{5} = \tfrac{61}{8}, \quad
c_{6} = \tfrac{1123}{1024} \pi^2-\tfrac{1039}{48}, \quad
c_{7} = \tfrac{1229711}{9600}-\tfrac{512}{5} \gamma +\tfrac{1139}{1024} \pi^2-\tfrac{1024}{5} \log(2), \quad
c_{7}^{{\rm ln}} = -\tfrac{256}{5},
\nonumber \\
c_{8} = -\tfrac{431650697}{403200}+\tfrac{14512}{105} \gamma +\tfrac{30981749}{196608} \pi^2+\tfrac{62128}{105} \log(2)-\tfrac{2187}{7} \log(3), \quad
c_{8}^{{\rm ln}} = \tfrac{7256}{105}, \quad
c_{8.5} = -\tfrac{54784}{525} \pi,
\nonumber \\
c_{9} = -\tfrac{165133169609}{14515200}+\tfrac{848360}{567} \gamma +\tfrac{14942182769}{14155776} \pi^2-\tfrac{33612139}{2097152} \pi^4-\tfrac{894008}{2835} \log(2)+\tfrac{33291}{14} \log(3),
\nonumber \\
c_{9}^{{\rm ln}} = \tfrac{424180}{567}, \quad
c_{9.5} = \tfrac{376087}{2205} \pi,
\end{gather*}
\begin{align}
c_{10} =& -\tfrac{8192}{5} \zeta(3)-\tfrac{26765202393708937}{391184640000}-\tfrac{38522675552}{5457375} \gamma +\tfrac{438272}{525} \gamma^2+\tfrac{395362037005919}{79272345600} \pi^2+\tfrac{109519423171}{536870912} \pi^4+\tfrac{1753088}{525} \log^2(2)
\nonumber \\
  & +\tfrac{14706174496}{5457375} \log(2)+\tfrac{1753088}{525} \gamma \log(2)-\tfrac{112748841}{24640} \log(3)-\tfrac{48828125}{19008} \log(5),
\nonumber \\
c_{10}^{{\rm ln}} =& -\tfrac{19261337776}{5457375}+\tfrac{438272}{525} \gamma +\tfrac{876544}{525} \log(2),
\nonumber \\
c_{10}^{{\rm ln}^2} =& \tfrac{109568}{525},
\nonumber \\
c_{10.5} =& \tfrac{3093926951}{2182950} \pi,
\nonumber \\
c_{11} =& -\tfrac{84064}{21} \zeta(3)+\tfrac{195654753392517961067}{155335864320000}-\tfrac{12929040365414}{496621125} \gamma -\tfrac{2313488}{11025} \gamma^2+\tfrac{2316722872897469}{123312537600} \pi^2-\tfrac{3620507818748053}{257698037760} \pi^4
\nonumber \\
  & -\tfrac{17798768}{1575} \log^2(2)+\tfrac{170586}{49} \log^2(3)-\tfrac{3889472783186}{55180125} \log(2)-\tfrac{86145568}{11025} \gamma \log(2)-\tfrac{43394903120133}{784784000} \log(3)
\nonumber \\
  & +\tfrac{341172}{49} \gamma \log(3)+\tfrac{341172}{49} \log(2) \log(3)+\tfrac{16724609375}{494208} \log(5),
\nonumber \\
c_{11}^{{\rm ln}} =& -\tfrac{6489947184307}{496621125}-\tfrac{2313488}{11025} \gamma -\tfrac{43072784}{11025} \log(2)+\tfrac{170586}{49} \log(3),
\nonumber \\
c_{11}^{{\rm ln}^2} =& -\tfrac{578372}{11025},
\nonumber \\
c_{11.5} =& -\tfrac{8576539777810427}{1048863816000} \pi +\tfrac{93790208}{55125} \gamma \pi -\tfrac{876544}{1575} \pi^3+\tfrac{187580416}{55125} \pi \log(2),
\nonumber \\
c_{11.5}^{{\rm ln}} =& \tfrac{46895104}{55125} \pi,
\nonumber \\
c_{12} =& \tfrac{76691344}{2835} \zeta(3)+\tfrac{895947046032111170223105769}{33493829736407040000}+\tfrac{336665071377836141}{3539915379000} \gamma -\tfrac{132146874856}{9823275} \gamma^2+\tfrac{30145651479528925511}{390654119116800} \pi^2
\nonumber \\
  & -\tfrac{115919631551153810401}{395824185999360} \pi^4+\tfrac{3602804277127}{3221225472} \pi^6+\tfrac{911509660696}{9823275} \log^2(2)-\tfrac{1298349}{49} \log^2(3)-\tfrac{500132500098693811}{3539915379000} \log(2)
\nonumber \\
  & +\tfrac{255763898416}{9823275} \gamma \log(2)+\tfrac{1361591585894889}{3139136000} \log(3)-\tfrac{2596698}{49} \gamma \log(3)-\tfrac{2596698}{49} \log(2) \log(3)
\nonumber \\
  & -\tfrac{20251009765625}{124540416} \log(5)-\tfrac{4747561509943}{277992000} \log(7),
\nonumber \\
c_{12}^{{\rm ln}} =& \tfrac{335124500204895341}{7079830758000}-\tfrac{132146874856}{9823275} \gamma +\tfrac{127881949208}{9823275} \log(2)-\tfrac{1298349}{49} \log(3),
\nonumber \\
c_{12}^{{\rm ln}^2} =& -\tfrac{33036718714}{9823275},
\nonumber \\
c_{12.5} =& -\tfrac{2772444986000367347}{136352296080000} \pi -\tfrac{58627948}{55125} \gamma \pi -\tfrac{4812988}{4725} \pi^3-\tfrac{368654676}{42875} \pi \log(2)+\tfrac{2217618}{343} \pi \log(3),
\nonumber \\
c_{12.5}^{{\rm ln}} =& -\tfrac{29313974}{55125} \pi,
\nonumber \\
c_{13} =& \tfrac{152706897076}{5457375} \zeta(3)+\tfrac{14024704}{525} \gamma \zeta(3)+\tfrac{131072}{5} \zeta(5)+\tfrac{28049408}{525} \zeta(3) \log(2)+\tfrac{45276263030368930088834341924361431}{148190770162354035916800000}
\nonumber \\
  & +\tfrac{775637589553645457793509}{3011937000222150000} \gamma +\tfrac{1965868177855822}{49165491375} \gamma^2-\tfrac{750321664}{165375} \gamma^3-\tfrac{588107630051246452676163101}{469253727883100160000} \pi^2+\tfrac{187580416}{33075} \gamma \pi^2
\nonumber \\
  & -\tfrac{34807846648736724821724433}{13299692649578496000} \pi^4-\tfrac{4209796222167113}{103079215104} \pi^6-\tfrac{6002573312}{165375} \log^3(2)-\tfrac{40943124194318002}{49165491375} \log^2(2)
\nonumber \\
  & -\tfrac{3001286656}{55125} \gamma \log^2(2)+\tfrac{125318296665}{2466464} \log^2(3)+\tfrac{188720703125}{4077216} \log^2(5)+\tfrac{9866911142625146186198093}{3011937000222150000} \log(2)
\nonumber \\
  & -\tfrac{15478057442549092}{49165491375} \gamma \log(2)-\tfrac{1500643328}{55125} \gamma^2 \log(2)+\tfrac{375160832}{33075} \pi^2 \log(2)-\tfrac{1600671398946305643}{1410817408000} \log(3)
\nonumber \\
  & +\tfrac{125318296665}{1233232} \gamma \log(3)+\tfrac{125318296665}{1233232} \log(2) \log(3)-\tfrac{75139342117568359375}{139874081098752} \log(5)+\tfrac{188720703125}{2038608} \gamma \log(5)
\nonumber \\
  & +\tfrac{188720703125}{2038608} \log(2) \log(5)+\tfrac{1488118310841113}{4725864000} \log(7),
\nonumber \\
c_{13}^{{\rm ln}} =& \tfrac{7012352}{525} \zeta(3)+\tfrac{768114287259147708033509}{6023874000444300000}+\tfrac{1965868177855822}{49165491375} \gamma -\tfrac{375160832}{55125} \gamma^2+\tfrac{93790208}{33075} \pi^2-\tfrac{1500643328}{55125} \log^2(2)
\nonumber \\
  & -\tfrac{7739028721274546}{49165491375} \log(2)-\tfrac{1500643328}{55125} \gamma \log(2)+\tfrac{125318296665}{2466464} \log(3)+\tfrac{188720703125}{4077216} \log(5),
\nonumber \\
c_{13}^{{\rm ln}^2} =& \tfrac{982934088927911}{98330982750}-\tfrac{187580416}{55125} \gamma -\tfrac{375160832}{55125} \log(2),
\nonumber \\
c_{13}^{{\rm ln}^3} =& -\tfrac{93790208}{165375},
\nonumber \\
c_{13.5} =& \tfrac{459787882287862634382511}{4448399261866560000} \pi -\tfrac{291464208472834}{11345882625} \gamma \pi +\tfrac{86968683554}{9823275} \pi^3+\tfrac{754663492020986}{34037647875} \pi \log(2)-\tfrac{16878537}{343} \pi \log(3),
\nonumber \\
c_{13.5}^{{\rm ln}} =& -\tfrac{145732104236417}{11345882625} \pi,
\nonumber \\
c_{14} =& -\tfrac{507112470513626}{496621125} \zeta(3)+\tfrac{1459707392}{11025} \gamma \zeta(3)+\tfrac{30330368}{105} \zeta(5)+\tfrac{31081984}{2205} \zeta(3) \log(2)+\tfrac{12282192}{49} \zeta(3) \log(3)
\nonumber \\
  & -\tfrac{20914536055020733601868714521492257200697}{7686655248321303843004416000000}-\tfrac{19314822847000337237144880769}{10415278146768194700000} \gamma +\tfrac{7162313664943370257}{22370298575625} \gamma^2-\tfrac{247666432}{42875} \gamma^3
\nonumber \\
  & +\tfrac{336606486091214661953625961098949}{3513771914388653998080000} \pi^2+\tfrac{61916608}{8575} \gamma \pi^2+\tfrac{455832122014651213374859602497}{142998295368267988992000} \pi^4+\tfrac{633213274864988114929}{415615395299328} \pi^6
\nonumber \\
  & +\tfrac{156132807424}{1157625} \log^3(2)-\tfrac{8870472}{343} \log^3(3)+\tfrac{85780419656473843313}{22370298575625} \log^2(2)+\tfrac{21030973696}{128625} \gamma \log^2(2)
\nonumber \\
  & -\tfrac{26611416}{343} \log^2(2) \log(3)+\tfrac{55342172390906997}{78556878400} \log^2(3)-\tfrac{26611416}{343} \gamma \log^2(3)-\tfrac{26611416}{343} \log(2) \log^2(3)
\nonumber \\
  & -\tfrac{64640615234375}{106007616} \log^2(5)-\tfrac{5166282262717961603753629417}{416611125870727788000} \log(2)+\tfrac{52416241905436677218}{22370298575625} \gamma \log(2)+\tfrac{3314595584}{77175} \gamma^2 \log(2)
\nonumber \\
  & -\tfrac{828648896}{46305} \pi^2 \log(2)-\tfrac{921591365191051244333481657}{162555171807508480000} \log(3)+\tfrac{55342172390906997}{39278439200} \gamma \log(3)-\tfrac{26611416}{343} \gamma^2 \log(3)
\nonumber \\
  & +\tfrac{11088090}{343} \pi^2 \log(3)+\tfrac{334116245930167881}{196392196000} \log(2) \log(3)-\tfrac{53222832}{343} \gamma \log(2) \log(3)+\tfrac{210377839241874220703125}{19742227446509568} \log(5)
\nonumber \\
  & -\tfrac{64640615234375}{53003808} \gamma \log(5)-\tfrac{64640615234375}{53003808} \log(2) \log(5)-\tfrac{1400878031194209247393}{597651664896000} \log(7),
\nonumber \\
c_{14}^{{\rm ln}} =& \tfrac{729853696}{11025} \zeta(3)-\tfrac{19443299901911886102163978369}{20830556293536389400000}+\tfrac{7162313664943370257}{22370298575625} \gamma -\tfrac{371499648}{42875} \gamma^2+\tfrac{30958304}{8575} \pi^2
\nonumber \\
  & +\tfrac{10515486848}{128625} \log^2(2)-\tfrac{13305708}{343} \log^2(3)+\tfrac{26226795763189455409}{22370298575625} \log(2)+\tfrac{3314595584}{77175} \gamma \log(2)
\nonumber \\
  & +\tfrac{55342172390906997}{78556878400} \log(3)-\tfrac{26611416}{343} \gamma \log(3)-\tfrac{26611416}{343} \log(2) \log(3)-\tfrac{64640615234375}{106007616} \log(5),
\nonumber \\
c_{14}^{{\rm ln}^2} =& \tfrac{7199663285885603857}{89481194302500}-\tfrac{185749824}{42875} \gamma +\tfrac{828648896}{77175} \log(2)-\tfrac{6652854}{343} \log(3),
\nonumber \\
c_{14}^{{\rm ln}^3} =& -\tfrac{30958304}{42875},
\nonumber \\
c_{14.5} =& \tfrac{1500643328}{55125} \pi \zeta(3)+\tfrac{12306435003814969601235891809399}{103875040717101461808000000} \pi +\tfrac{673907554336490184031}{6561954248850000} \gamma \pi -\tfrac{80284418048}{5788125} \gamma^2 \pi 
\nonumber \\
  & +\tfrac{18873449053136929}{3933239310000} \pi^3+\tfrac{1500643328}{165375} \gamma \pi^3+\tfrac{14024704}{23625} \pi^5-\tfrac{321137672192}{5788125} \pi \log^2(2)-\tfrac{40135482632184728921851}{177172764718950000} \pi \log(2)
\nonumber \\
  & -\tfrac{321137672192}{5788125} \gamma \pi \log(2)+\tfrac{3001286656}{165375} \pi^3 \log(2)+\tfrac{29046103665558867}{308616308000} \pi \log(3)+\tfrac{145881103515625}{1749125664} \pi \log(5),
\nonumber \\
c_{14.5}^{{\rm ln}} =& \tfrac{673907554336490184031}{13123908497700000} \pi -\tfrac{80284418048}{5788125} \gamma \pi +\tfrac{750321664}{165375} \pi^3-\tfrac{160568836096}{5788125} \pi \log(2),
\nonumber \\
c_{14.5}^{{\rm ln}^2} =& -\tfrac{20071104512}{5788125} \pi,
\nonumber \\
c_{15} =& -\tfrac{19017674025842159417}{5309873068500} \zeta(3)-\tfrac{1986297949952}{9823275} \gamma \zeta(3)+\tfrac{394576384}{567} \zeta(5)+\tfrac{31637350502656}{9823275} \zeta(3) \log(2)-\tfrac{93481128}{49} \zeta(3) \log(3)
\nonumber \\
  & -\tfrac{2156981124861492206459973244667903910764345212519}{15834448318299899346158352924672000000}-\tfrac{2239134931878759971610842802204107}{315520436178195690241800000} \gamma
\nonumber \\
  &  -\tfrac{525646348187213988152273}{2710743300199935000} \gamma^2+\tfrac{7068347307704576}{102112943625} \gamma^3+\tfrac{8886137056252424329336470110571506819}{2419231963056588277678080000} \pi^2-\tfrac{1767086826926144}{20422588725} \gamma \pi^2
\nonumber \\
  & +\tfrac{651241387862373001412589433250920937}{1359055799180018967379968000} \pi^4+\tfrac{12070914117030147676265915}{212795082393255936} \pi^6-\tfrac{2547624607151695403}{19791209299968} \pi^8
\nonumber \\
  & -\tfrac{335912485548351232}{102112943625} \log^3(2)+\tfrac{67514148}{343} \log^3(3)+\tfrac{3486208535559431080409581}{246431209109085000} \log^2(2)-\tfrac{8934542785182976}{3094331625} \gamma \log^2(2)
\nonumber \\
  & +\tfrac{202542444}{343} \log^2(2) \log(3)-\tfrac{811932090874103080377}{133546693280000} \log^2(3)+\tfrac{202542444}{343} \gamma \log^2(3)+\tfrac{202542444}{343} \log(2) \log^2(3)
\nonumber \\
  & +\tfrac{3103131723564453125}{1059652129536} \log^2(5)+\tfrac{99266763611398187}{230385870000} \log^2(7)-\tfrac{181440728366064415737944713721311}{6310408723563913804836000} \log(2)
\nonumber \\
  & +\tfrac{1496178473752627794534563}{271074330019993500} \gamma \log(2)-\tfrac{19214295857045248}{34037647875} \gamma^2 \log(2)+\tfrac{4803573964261312}{20422588725} \pi^2 \log(2)
\nonumber \\
  & +\tfrac{2029186475052690112702362124587}{27634379207276441600000} \log(3)-\tfrac{811932090874103080377}{66773346640000} \gamma \log(3)+\tfrac{202542444}{343} \gamma^2 \log(3)-\tfrac{84392685}{343} \pi^2 \log(3)
\nonumber \\
  & -\tfrac{1120809504525849732537}{66773346640000} \log(2) \log(3)+\tfrac{405084888}{343} \gamma \log(2) \log(3)-\tfrac{6883343827424302422607421875}{140217611841930534912} \log(5)
\nonumber \\
  & +\tfrac{3103131723564453125}{529826064768} \gamma \log(5)+\tfrac{3103131723564453125}{529826064768} \log(2) \log(5)-\tfrac{1229477788055969228362223323}{1176841869863362560000} \log(7)
\nonumber \\
  & +\tfrac{99266763611398187}{115192935000} \gamma \log(7)+\tfrac{99266763611398187}{115192935000} \log(2) \log(7),
\nonumber \\
c_{15}^{{\rm ln}} =& -\tfrac{993148974976}{9823275} \zeta(3)-\tfrac{2255624992466899593198995408162507}{631040872356391380483600000}-\tfrac{516171613631379164677073}{2710743300199935000} \gamma +\tfrac{3534173653852288}{34037647875} \gamma^2
\nonumber \\
  & -\tfrac{883543413463072}{20422588725} \pi^2-\tfrac{4467271392591488}{3094331625} \log^2(2)+\tfrac{101271222}{343} \log^2(3)+\tfrac{1501976725789824367728803}{542148660039987000} \log(2)
\nonumber \\
  & -\tfrac{19214295857045248}{34037647875} \gamma \log(2)-\tfrac{811932090874103080377}{133546693280000} \log(3)+\tfrac{202542444}{343} \gamma \log(3)+\tfrac{202542444}{343} \log(2) \log(3)
\nonumber \\
  & +\tfrac{3103131723564453125}{1059652129536} \log(5)+\tfrac{99266763611398187}{230385870000} \log(7),
\nonumber \\
c_{15}^{{\rm ln}^2} =& -\tfrac{486613296926917903160273}{10842973200799740000}+\tfrac{1767086826926144}{34037647875} \gamma -\tfrac{4803573964261312}{34037647875} \log(2)+\tfrac{50635611}{343} \log(3),
\nonumber \\
c_{15}^{{\rm ln}^3} =& \tfrac{883543413463072}{102112943625},
\nonumber \\
c_{15.5} =& \tfrac{8660541616}{77175} \pi \zeta(3)-\tfrac{15968906809329459860858879418922575689}{9127903107381548313567744000000} \pi +\tfrac{17429771806026061957846879}{32245443178848900000} \gamma \pi -\tfrac{258366416648}{24310125} \gamma^2 \pi 
\nonumber \\
  & -\tfrac{72458812852556068207}{238616518140000} \pi^3+\tfrac{8660541616}{231525} \gamma \pi^3+\tfrac{568239632}{99225} \pi^5+\tfrac{21107520552152}{121550625} \pi \log^2(2)-\tfrac{172974204}{2401} \pi \log^2(3)
\nonumber \\
  & +\tfrac{63856957049461045127565151}{32245443178848900000} \pi \log(2)+\tfrac{12349237590448}{121550625} \gamma \pi \log(2)-\tfrac{1072411696}{385875} \pi^3 \log(2)
\nonumber \\
  & +\tfrac{2618534544249464439579}{1965885881960000} \pi \log(3)-\tfrac{345948408}{2401} \gamma \pi \log(3)+\tfrac{26611416}{343} \pi^3 \log(3)-\tfrac{345948408}{2401} \pi \log(2) \log(3)
\nonumber \\
  & -\tfrac{49967195576171875}{45477267264} \pi \log(5),
\nonumber \\
c_{15.5}^{{\rm ln}} =& \tfrac{17429771806026061957846879}{64490886357697800000} \pi -\tfrac{258366416648}{24310125} \gamma \pi +\tfrac{4330270808}{231525} \pi^3+\tfrac{6174618795224}{121550625} \pi \log(2)-\tfrac{172974204}{2401} \pi \log(3),
\nonumber \\
c_{15.5}^{{\rm ln}^2} =& -\tfrac{64591604162}{24310125} \pi.
\end{align}

\section{Post-Newtonian expansion coefficients for \texorpdfstring{$\Delta \lambda^B$}{Delta lambda B}}
\label{sec:DeltaLambdaB}
In this Appendix, we give the analytic post-Newtonian expansion coefficients
for the tidal invariant $\Delta \lambda^B$ (see \cite{Dolan:2014pja} for a definition) up to order $y^{15.5}$. We have also computed
higher order coefficients up to order $y^{21.5}$, but they are too
long to give here; instead, we make them available electronically \cite{online}.
\begin{gather*}
c_{3.5} = 2, \quad
c_{4.5} = 3, \quad
c_{5.5} = \tfrac{59}{4}, \quad
c_{6.5} = \tfrac{2761}{24}-\tfrac{41}{16} \pi^2, \quad
c_{7.5} = \tfrac{1618039}{2880}+\tfrac{1808}{15} \gamma-\tfrac{112919}{3072} \pi^2+240 \log(2), \quad
c_{7.5}^{{\rm ln}} = \tfrac{904}{15},
\nonumber \\
c_{8.5} = \tfrac{491047651}{201600}+\tfrac{2756}{105} \gamma-\tfrac{565685}{3072} \pi^2-\tfrac{4492}{21} \log(2)+\tfrac{3645}{14} \log(3), \quad
c_{8.5}^{{\rm ln}} = \tfrac{1378}{105}, \quad
c_{9} = \tfrac{856}{7} \pi,
\nonumber \\
c_{9.5} = -\tfrac{454873888681}{50803200}-\tfrac{3881396}{2835} \gamma+\tfrac{7377893735}{3538944} \pi^2-\tfrac{26691349}{524288} \pi^4-\tfrac{1992212}{2835} \log(2)-\tfrac{200961}{140} \log(3), \quad
c_{9.5}^{{\rm ln}} = -\tfrac{1940698}{2835},
\end{gather*}
\begin{align}
\nonumber \\
c_{10} =& -\tfrac{69473}{22050} \pi,
\nonumber \\
c_{10.5} =& \tfrac{28736}{15} \zeta(3)-\tfrac{423235951437871681}{1760330880000}+\tfrac{83360241649}{10914750} \gamma-\tfrac{1537376}{1575} \gamma^2+\tfrac{42496203125923}{2477260800} \pi^2+\tfrac{89531499967}{100663296} \pi^4-\tfrac{6139232}{1575} \log^2(2)
\nonumber \\
  & +\tfrac{82889847697}{10914750} \log(2)-\tfrac{2047552}{525} \gamma \log(2)+\tfrac{1412559}{1232} \log(3)+\tfrac{9765625}{7128} \log(5),
\nonumber \\
c_{10.5}^{{\rm ln}} =& \tfrac{83360241649}{21829500}-\tfrac{1537376}{1575} \gamma-\tfrac{1023776}{525} \log(2),
\nonumber \\
c_{10.5}^{{\rm ln}^2} =& -\tfrac{384344}{1575},
\nonumber \\
c_{11} =& -\tfrac{5843221973}{4365900} \pi,
\nonumber \\
c_{11.5} =& \tfrac{583232}{105} \zeta(3)-\tfrac{21309701350511477421907}{9397819791360000}+\tfrac{42734893209857}{1324323000} \gamma-\tfrac{10248496}{11025} \gamma^2+\tfrac{41783113230548567}{554906419200} \pi^2+\tfrac{995995846414723}{64424509440} \pi^4
\nonumber \\
  & +\tfrac{55366352}{11025} \log^2(2)-\tfrac{142155}{49} \log^2(3)+\tfrac{66996912132433}{1324323000} \log(2)+\tfrac{4649248}{2205} \gamma \log(2)+\tfrac{8818395978069}{196196000} \log(3)
\nonumber \\
  & -\tfrac{284310}{49} \gamma \log(3)-\tfrac{284310}{49} \log(2) \log(3)-\tfrac{5041015625}{324324} \log(5),
\nonumber \\
c_{11.5}^{{\rm ln}} =& \tfrac{42972211891457}{2648646000}-\tfrac{10248496}{11025} \gamma+\tfrac{2324624}{2205} \log(2)-\tfrac{142155}{49} \log(3),
\nonumber \\
c_{11.5}^{{\rm ln}^2} =& -\tfrac{2562124}{11025},
\nonumber \\
c_{12} =& \tfrac{177773734055963}{20170458000} \pi-\tfrac{109544032}{55125} \gamma \pi+\tfrac{1023776}{1575} \pi^3-\tfrac{656897824}{165375} \pi \log(2),
\nonumber \\
c_{12}^{{\rm ln}} =& -\tfrac{54772016}{55125} \pi,
\nonumber \\
c_{12.5} =& -\tfrac{61736408}{2835} \zeta(3)+\tfrac{650842678174698441560847239}{31101413326663680000}-\tfrac{1009884112621826107}{14159661516000} \gamma+\tfrac{115048067276}{9823275} \gamma^2+\tfrac{2612085785441191481}{6975966412800} \pi^2
\nonumber \\
  & -\tfrac{14375375047759450747}{49478023249920} \pi^4+\tfrac{1594697066963}{402653184} \pi^6-\tfrac{2526608396}{56133} \log^2(2)+\tfrac{433572021}{26950} \log^2(3)+\tfrac{1647998244938492453}{14159661516000} \log(2)
\nonumber \\
  & -\tfrac{27159862376}{9823275} \gamma \log(2)-\tfrac{496811505579201}{2158156000} \log(3)+\tfrac{433572021}{13475} \gamma \log(3)+\tfrac{433572021}{13475} \log(2) \log(3)+\tfrac{288888671875}{4790016} \log(5)
\nonumber \\
  & +\tfrac{678223072849}{101088000} \log(7),
\nonumber \\
c_{12.5}^{{\rm ln}} =& -\tfrac{993028451553179707}{28319323032000}+\tfrac{115048067276}{9823275} \gamma-\tfrac{13579931188}{9823275} \log(2)+\tfrac{433572021}{26950} \log(3),
\nonumber \\
c_{12.5}^{{\rm ln}^2} =& \tfrac{28762016819}{9823275},
\nonumber \\
c_{13} =& \tfrac{5157596639147063}{180360180000} \pi-\tfrac{14606534}{11025} \gamma \pi+\tfrac{1495618}{945} \pi^3+\tfrac{1061371118}{385875} \pi \log(2)-\tfrac{1848015}{343} \pi \log(3),
\nonumber \\
c_{13}^{{\rm ln}} =& -\tfrac{7303267}{11025} \pi,
\nonumber \\
c_{13.5} =& \tfrac{6520649816}{202125} \zeta(3)-\tfrac{49113856}{1575} \gamma \zeta(3)-\tfrac{459008}{15} \zeta(5)-\tfrac{6546688}{105} \zeta(3) \log(2)+\tfrac{137738077427722761383939335398968819}{222286155243531053875200000}
\nonumber \\
  & -\tfrac{192574199072389690922867}{3011937000222150000} \gamma-\tfrac{2572458931724062}{49165491375} \gamma^2+\tfrac{2627591296}{496125} \gamma^3-\tfrac{3242997919731208736780905997}{1055820887736975360000} \pi^2-\tfrac{656897824}{99225} \gamma \pi^2
\nonumber \\
  & -\tfrac{51226131330048617152601737}{9974769487183872000} \pi^4-\tfrac{148094221748880223}{1623497637888} \pi^6+\tfrac{259388544}{6125} \log^3(2)+\tfrac{506021443260742}{1820944125} \log^2(2)
\nonumber \\
  & +\tfrac{10505968768}{165375} \gamma \log^2(2)-\tfrac{203748051807}{15415400} \log^2(3)-\tfrac{37744140625}{1528956} \log^2(5)-\tfrac{935831561675894643798323}{602387400044430000} \log(2)
\nonumber \\
  & +\tfrac{31405756367204}{1966619655} \gamma \log(2)+\tfrac{350247808}{11025} \gamma^2 \log(2)-\tfrac{87561952}{6615} \pi^2 \log(2)+\tfrac{1517581031980000347}{5246477236000} \log(3)
\nonumber \\
  & -\tfrac{203748051807}{7707700} \gamma \log(3)-\tfrac{203748051807}{7707700} \log(2) \log(3)+\tfrac{79529650673427734375}{209811121648128} \log(5)-\tfrac{37744140625}{764478} \gamma \log(5)
\nonumber \\
  & -\tfrac{37744140625}{764478} \log(2) \log(5)-\tfrac{12722592845553577}{113420736000} \log(7),
\nonumber \\
c_{13.5}^{{\rm ln}} =& -\tfrac{24556928}{1575} \zeta(3)-\tfrac{172840217327848466762867}{6023874000444300000}-\tfrac{2572458931724062}{49165491375} \gamma+\tfrac{1313795648}{165375} \gamma^2-\tfrac{328448912}{99225} \pi^2+\tfrac{5252984384}{165375} \log^2(2)
\nonumber \\
  & +\tfrac{15702878183602}{1966619655} \log(2)+\tfrac{350247808}{11025} \gamma \log(2)-\tfrac{203748051807}{15415400} \log(3)-\tfrac{37744140625}{1528956} \log(5),
\nonumber \\
c_{13.5}^{{\rm ln}^2} =& -\tfrac{1286229465862031}{98330982750}+\tfrac{656897824}{165375} \gamma+\tfrac{87561952}{11025} \log(2),
\nonumber \\
c_{13.5}^{{\rm ln}^3} =& \tfrac{328448912}{496125},
\nonumber \\
c_{14} =& -\tfrac{182312652332925353457079}{2224199630933280000} \pi+\tfrac{261156926572103}{11345882625} \gamma \pi-\tfrac{72277933843}{9823275} \pi^3+\tfrac{7448963901653}{34037647875} \pi \log(2)+\tfrac{311318752347}{10375750} \pi \log(3),
\nonumber \\
c_{14}^{{\rm ln}} =& \tfrac{261156926572103}{22691765250} \pi,
\nonumber \\
c_{14.5} =& \tfrac{226437359033}{210210} \zeta(3)-\tfrac{229157312}{1575} \gamma \zeta(3)-\tfrac{4140608}{15} \zeta(5)-\tfrac{904209472}{11025} \zeta(3) \log(2)-\tfrac{10235160}{49} \zeta(3) \log(3)
\nonumber \\
  & +\tfrac{8717230186879021394846309680393737429919}{1333399379810838421745664000000}+\tfrac{26420013359615317979764911787}{11903175024877936800000} \gamma-\tfrac{4548539361797409137}{12783027757500} \gamma^2+\tfrac{1769521504}{165375} \gamma^3
\nonumber \\
  & -\tfrac{341456877370199774633224097417}{3431417885145169920000} \pi^2-\tfrac{442380376}{33075} \gamma \pi^2-\tfrac{2563963105871968704720413161}{63838524717976780800} \pi^4-\tfrac{11091687187004415235}{6493990551552} \pi^6
\nonumber \\
  & -\tfrac{2162255392}{33075} \log^3(2)+\tfrac{7392060}{343} \log^3(3)-\tfrac{155965565348317143767}{89481194302500} \log^2(2)-\tfrac{25355744864}{385875} \gamma \log^2(2)
\nonumber \\
  & +\tfrac{22176180}{343} \log^2(2) \log(3)-\tfrac{53641337092064217}{98196098000} \log^2(3)+\tfrac{22176180}{343} \gamma \log^2(3)+\tfrac{22176180}{343} \log(2) \log^2(3)
\nonumber \\
  & +\tfrac{15194814453125}{54108054} \log^2(5)+\tfrac{62687419667570338818465670693}{9258025019349506400000} \log(2)-\tfrac{829912983765218981}{596541295350} \gamma \log(2)-\tfrac{194147936}{385875} \gamma^2 \log(2)
\nonumber \\
  & +\tfrac{48536984}{231525} \pi^2 \log(2)+\tfrac{370647543893739245573111253}{81277585903754240000} \log(3)-\tfrac{53641337092064217}{49098049000} \gamma \log(3)+\tfrac{22176180}{343} \gamma^2 \log(3)
\nonumber \\
  & -\tfrac{9240075}{343} \pi^2 \log(3)-\tfrac{12026722532147397}{9819609800} \log(2) \log(3)+\tfrac{44352360}{343} \gamma \log(2) \log(3)-\tfrac{4794041153062144818359375}{967369144878968832} \log(5)
\nonumber \\
  & +\tfrac{15194814453125}{27054027} \gamma \log(5)+\tfrac{15194814453125}{27054027} \log(2) \log(5)+\tfrac{661364683359445043153}{896477497344000} \log(7),
\nonumber \\
c_{14.5}^{{\rm ln}} =& -\tfrac{114578656}{1575} \zeta(3)+\tfrac{26816721352469623226670102187}{23806350049755873600000}-\tfrac{4548539361797409137}{12783027757500} \gamma+\tfrac{884760752}{55125} \gamma^2-\tfrac{221190188}{33075} \pi^2
\nonumber \\
  & -\tfrac{12677872432}{385875} \log^2(2)+\tfrac{11088090}{343} \log^2(3)-\tfrac{830908973657011877}{1193082590700} \log(2)-\tfrac{194147936}{385875} \gamma \log(2)
\nonumber \\
  & -\tfrac{53641337092064217}{98196098000} \log(3)+\tfrac{22176180}{343} \gamma \log(3)+\tfrac{22176180}{343} \log(2) \log(3)+\tfrac{15194814453125}{54108054} \log(5),
\nonumber \\
c_{14.5}^{{\rm ln}^2} =& -\tfrac{4569882002335828337}{51132111030000}+\tfrac{442380376}{55125} \gamma-\tfrac{48536984}{385875} \log(2)+\tfrac{5544045}{343} \log(3),
\nonumber \\
c_{14.5}^{{\rm ln}^3} =& \tfrac{221190188}{165375},
\nonumber \\
c_{15} =& -\tfrac{350247808}{11025} \pi \zeta(3)+\tfrac{5339130148856440616477543938057}{93487536645391315627200000} \pi-\tfrac{374731983268181645111}{2952879411982500} \gamma \pi+\tfrac{18738257728}{1157625} \gamma^2 \pi
\nonumber \\
  & +\tfrac{229453971195409}{17878360500} \pi^3-\tfrac{350247808}{33075} \gamma \pi^3-\tfrac{3273344}{4725} \pi^5+\tfrac{13877287104}{214375} \pi \log^2(2)-\tfrac{2139817700265597128761}{44293191179737500} \pi \log(2)
\nonumber \\
  & +\tfrac{1124138658176}{17364375} \gamma \pi \log(2)-\tfrac{10505968768}{496125} \pi^3 \log(2)-\tfrac{1940642256399057}{77154077000} \pi \log(3)-\tfrac{29176220703125}{655922124} \pi \log(5),
\nonumber \\
c_{15}^{{\rm ln}} =& -\tfrac{374731983268181645111}{5905758823965000} \pi+\tfrac{18738257728}{1157625} \gamma \pi-\tfrac{175123904}{33075} \pi^3+\tfrac{562069329088}{17364375} \pi \log(2),
\nonumber \\
c_{15}^{{\rm ln}^2} =& \tfrac{4684564432}{1157625} \pi,
\nonumber \\
c_{15.5} =& \tfrac{6202424414638085951}{2654936534250} \zeta(3)+\tfrac{1780141198336}{9823275} \gamma \zeta(3)-\tfrac{1316562304}{2835} \zeta(5)-\tfrac{18782726706176}{9823275} \zeta(3) \log(2)
\nonumber \\
  & +\tfrac{15608592756}{13475} \zeta(3) \log(3)-\tfrac{13274137299499961392016982771705466784182009781147}{150427259023849043788504352784384000000}+\tfrac{199145605807824782900827181923163}{40877141529158955820800000} \gamma
\nonumber \\
  & +\tfrac{42790957420641639646681}{208518715399995000} \gamma^2-\tfrac{6046888630600768}{102112943625} \gamma^3+\tfrac{14216496541215185765925039372537482329}{4838463926113176555356160000} \pi^2+\tfrac{1511722157650192}{20422588725} \gamma \pi^2
\nonumber \\
  & +\tfrac{85771577477966663370703907387338853}{339763949795004741844992000} \pi^4+\tfrac{2170911676656317357689133}{53198770598313984} \pi^6-\tfrac{2392749397956998243}{4947802324992} \pi^8
\nonumber \\
  & +\tfrac{40210046347331776}{20422588725} \log^3(2)-\tfrac{622637504694}{5187875} \log^3(3)-\tfrac{32926195046864358119449531}{2710743300199935000} \log^2(2)+\tfrac{55640766180961216}{34037647875} \gamma \log^2(2)
\nonumber \\
  & -\tfrac{1867912514082}{5187875} \log^2(2) \log(3)+\tfrac{456376186270608529383}{146901362608000} \log^2(3)-\tfrac{1867912514082}{5187875} \gamma \log^2(3)
\nonumber \\
  & -\tfrac{1867912514082}{5187875} \log(2) \log^2(3)-\tfrac{96693640009765625}{88304344128} \log^2(5)-\tfrac{14180966230199741}{83776680000} \log^2(7)
\nonumber \\
  & +\tfrac{14195853596778147050357617416713711}{531402839879066425670400000} \log(2)-\tfrac{6903734693281690276174139}{1355371650099967500} \gamma \log(2)+\tfrac{8773912142318528}{34037647875} \gamma^2 \log(2)
\nonumber \\
  & -\tfrac{2193478035579632}{20422588725} \pi^2 \log(2)-\tfrac{5014469439425452168758965467887}{151989085640020428800000} \log(3)+\tfrac{456376186270608529383}{73450681304000} \gamma \log(3)
\nonumber \\
  & -\tfrac{1867912514082}{5187875} \gamma^2 \log(3)+\tfrac{311318752347}{2075150} \pi^2 \log(3)+\tfrac{2966133302205799233411}{367253406520000} \log(2) \log(3)
\nonumber \\
  & -\tfrac{3735825028164}{5187875} \gamma \log(2) \log(3)+\tfrac{47489036363197548687095703125}{2664134624996680163328} \log(5)-\tfrac{96693640009765625}{44152172064} \gamma \log(5)
\nonumber \\
  & -\tfrac{96693640009765625}{44152172064} \log(2) \log(5)+\tfrac{82461965144461117073077709}{53492812266516480000} \log(7)-\tfrac{14180966230199741}{41888340000} \gamma \log(7)
\nonumber \\
  & -\tfrac{14180966230199741}{41888340000} \log(2) \log(7),
\nonumber \\
c_{15.5}^{{\rm ln}} =& \tfrac{890070599168}{9823275} \zeta(3)+\tfrac{205456888221800274077211649430363}{81754283058317911641600000}+\tfrac{42097718879672354936281}{208518715399995000} \gamma-\tfrac{3023444315300384}{34037647875} \gamma^2
\nonumber \\
  & +\tfrac{755861078825096}{20422588725} \pi^2+\tfrac{27820383090480608}{34037647875} \log^2(2)-\tfrac{933956257041}{5187875} \log^2(3)-\tfrac{6938515922161787047246139}{2710743300199935000} \log(2)
\nonumber \\
  & +\tfrac{8773912142318528}{34037647875} \gamma \log(2)+\tfrac{456376186270608529383}{146901362608000} \log(3)-\tfrac{1867912514082}{5187875} \gamma \log(3)-\tfrac{1867912514082}{5187875} \log(2) \log(3)
\nonumber \\
  & -\tfrac{96693640009765625}{88304344128} \log(5)-\tfrac{14180966230199741}{83776680000} \log(7),
\nonumber \\
c_{15.5}^{{\rm ln}^2} =& \tfrac{38836915123602398840281}{834074861599980000}-\tfrac{1511722157650192}{34037647875} \gamma+\tfrac{2193478035579632}{34037647875} \log(2)-\tfrac{933956257041}{10375750} \log(3),
\nonumber \\
c_{15.5}^{{\rm ln}^3} =& -\tfrac{755861078825096}{102112943625}.
\end{align}

\end{widetext}

\bibliographystyle{apsrev4-1}
\bibliography{MST-PN}

\end{document}